\documentclass[prr,reprint,twocolumn,footinbib,longbibliography,superscriptaddress]{revtex4-2}
\usepackage{amsmath}
\usepackage{amssymb}
\usepackage{amsfonts}
\usepackage{times}
\usepackage[dvipdfmx]{graphicx}
\usepackage[usenames,dvipsnames]{xcolor}
\usepackage{bm}
\usepackage{amsthm}
\usepackage{comment}

\usepackage{physics}
\usepackage{mathrsfs}

\usepackage{hyperref}

\usepackage{empheq} 

\newcommand{\tp}{\tilde{p}} 
\newcommand{\td}{\tilde{\delta}} 
\newcommand{\tb}{\tilde{\alpha}} 

\begin{document}

\title{Game-theoretical approach to minimum entropy productions in information thermodynamics}
\date{\today}

\author{Yuma Fujimoto}
\affiliation{Department of Evolutionary Studies of Biosystems, SOKENDAI, Shonan Village, Hayama, Kanagawa 240-0193, Japan}
\affiliation{Universal Biology Institute (UBI), the University of Tokyo, 7-3-1 Hongo, Bunkyo-ku, Tokyo 113-0033, Japan}
\affiliation{CyberAgent, AI Lab, Shibuya-ku 150-0042, Japan}
\author{Sosuke Ito}
\affiliation{Universal Biology Institute (UBI), the University of Tokyo, 7-3-1 Hongo, Bunkyo-ku, Tokyo 113-0033, Japan}
\affiliation{JST, PRESTO, 4-1-8 Honcho, Kawaguchi, Saitama, 332-0012, Japan}

\begin{abstract}
In a situation where each player has control over the transition probabilities of each subsystem, we game-theoretically analyze the optimization problem of minimizing both the partial entropy production of each subsystem and a penalty for failing to achieve a given
state transition. In the regime of linear irreversible thermodynamics, we obtain the Nash equilibrium solution of the probability flow and calculate each partial entropy production for this solution. We find a trade-off such that a partial entropy production should be larger if we want the other partial entropy production to be smaller. The total entropy production can be minimized if each subsystem equally shares the penalty. We identify that this trade-off is due to the interacting contribution of the probability flow and discuss a possible biological validity for Escherichia coli chemotaxis.
\end{abstract}
\maketitle

\section{Introduction}
In physical systems such as living systems, there universally exist situations where many-body systems cooperatively perform a state transition to achieve a given task. Achievement of a given task generally takes a physical cost. When the state transition is performed by physical stochastic processes~\cite{van1992stochastic}, the cost can be introduced as the entropy production in stochastic thermodynamics~\cite{schnakenberg1976network,sekimoto2010stochastic,seifert2012stochastic}. Minimizing this entropy production might be crucial for the many-body systems to maintain their performance.

The minimization problem of entropy production has been known in the context of optimal control in stochastic thermodynamics. Historically, this minimization problem was discussed to improve the efficiency of a stochastic heat engine and the work extraction~\cite{van2005thermodynamic,schmiedl2007optimal,schmiedl2007efficiency,maillet2019optimal}. Recent progress of stochastic thermodynamics clarified that such a minimization problem is related to differential geometry~\cite{aurell2011optimal,aurell2012refined,sivak2012thermodynamic,maes2017frenetic,ito2018stochastic,ito2020stochastic,dechant2019thermodynamic,proesmans2020finite,proesmans2020optimal, van2021geometrical,nakazato2021geometrical, dechant2022geometric,dechant2022minimum,hamazaki2022speed,yoshimura2023housekeeping}, such as the optimal transport theory~\cite{villani2009optimal,benamou2000computational, villani2021topics,otto2000generalization,arnold2001convex,villani2008entropy, maas2011gradient,cuturi2013sinkhorn} and information geometry~\cite{amari2000methods, crooks2007measuring, amari2016information, ito2020unified}. This minimization problem has not been frequently discussed for many-body systems because we consider the total entropy production for a single whole system in a stochastic heat engine.

For many-body systems, the entropy production for the whole system can be partitioned into partial entropy productions of the subsystems~\cite{allahverdyan2009thermodynamic,ito2013information,horowitz2014thermodynamics,hartich2014stochastic,horowitz2014second, shiraishi2015fluctuation, ito2015maxwell, shiraishi2015role, ito2016backward, spinney2016transfer, ito2020unified, wolpert2020uncertainty, nakazato2021geometrical}. This partitioning is originally introduced for non-autonomous system in information thermodynamics~\cite{still2012thermodynamics, sagawa2012fluctuation, parrondo2015thermodynamics} and is applied to autonomous information processing in living systems~\cite{barato2014efficiency, sartori2014thermodynamic, ito2015maxwell,bo2015thermodynamic, hartich2016sensory, ouldridge2017thermodynamics, mcgrath2017biochemical,auconi2019information, skinner2021improved,yoshida2022thermodynamic}. Although the minimization problem of the partial entropy production is important to achieve an efficient information-to-free energy conversion~\cite{sagawa2010generalized, toyabe2010experimental, koski2015chip, paneru2018optimal, manzano2021thermodynamics}, a situation of the two-body system is often seen that the one partial entropy production should be increased to reduce the other partial entropy production in a living system. The total entropy production for the whole system is not minimized in this situation, and this situation is not well treated in the conventional minimization problem of the total entropy production. Therefore, a framework has been required to deal with a certain kind of thermodynamic optimality when many-body systems cooperatively achieve a given task. Such optimality is often discussed in the field of game theory \cite{nash1951non, von2007theory,osborne1994course,myerson1997game,press2012iterated, fujimoto2019emergence,fujimoto2021exploitation} or the mean-field game theory~\cite{jovanovic1988anonymous,lasry2007mean,bensoussan2013mean,gomes2016regularity,ruthotto2020machine}. For example, the prisoner's dilemma game is used to analyze a cooperative behavior of multi-agent systems in game theory~\cite{axelrod1981evolution}, and the Nash equilibrium~\cite{nash1950equilibrium} gives the optimality for multi-agent systems.

In this study, we newly introduce a game-theoretic framework for a conflict between minimizations of partial entropy productions of two subsystems $X$ and $Y$ with a given task on the whole system. We treat this optimization problem with the Markov jump process for the bipartite system under the near-equilibrium condition and introduce the concept of Nash equilibrium into stochastic thermodynamics to explain this conflict. This Nash equilibrium solution shows a trade-off such that the partial entropy production of $X$ should be increased when the partial entropy production of $Y$ is reduced. Moreover, this conflict provides an inevitable dissipation in the Nash equilibrium solution, and this dissipation is expressed by the interacting contribution of the probability flow. We illustrate these facts by the numerical calculations and discuss a possible biological validity for {\it Escherichia coli} ({\it E. coli}) chemotaxis in terms of evolutionary processes. We also remark on our game-theoretic minimization of partial entropy productions in terms of optimal transport theory.

\section{Setup}
\begin{figure*}
    \centering
    \includegraphics[width=0.7\hsize]{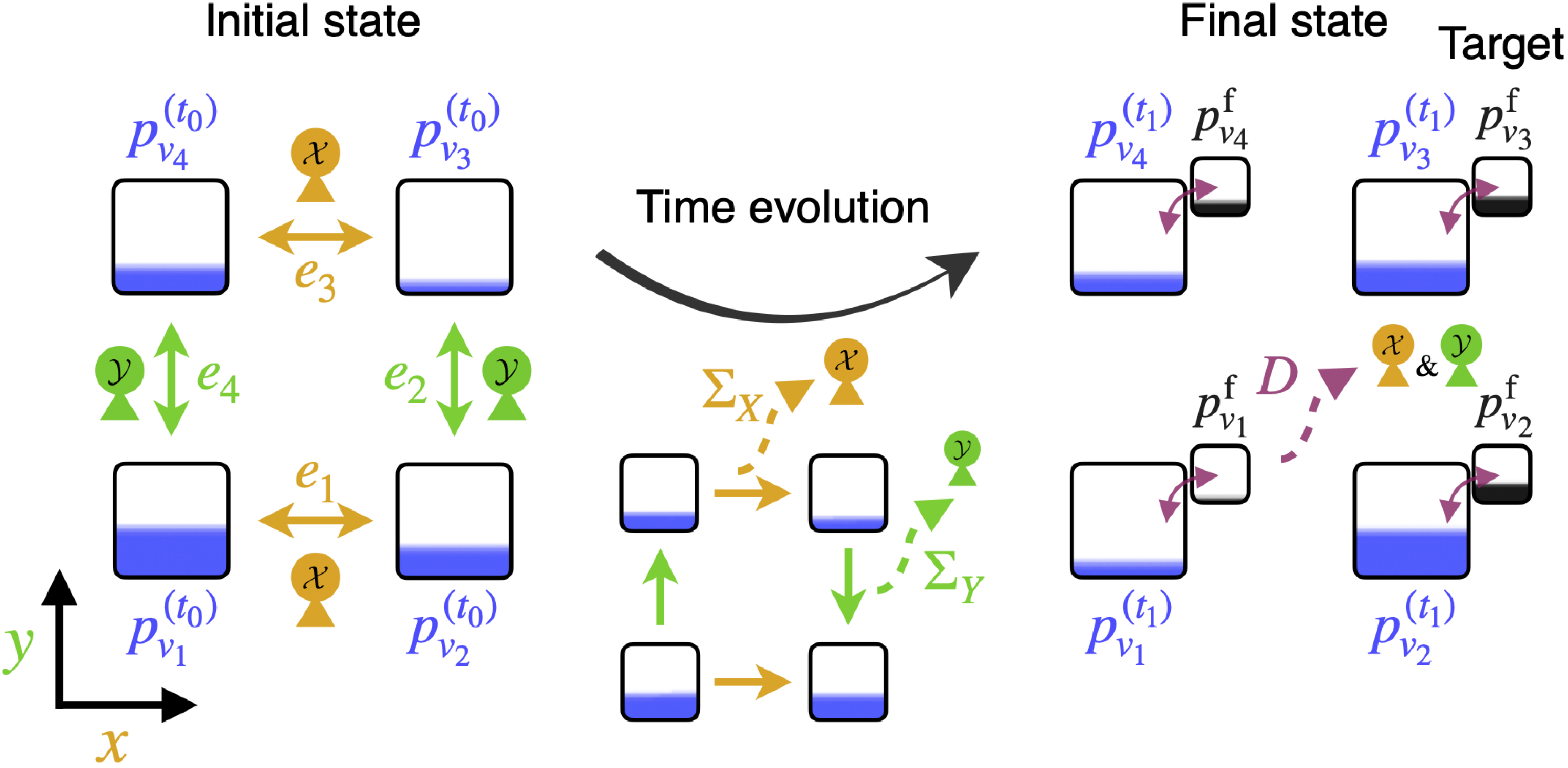}
    \caption{Schematics of settings. Blue bars indicate a probability distribution on space $\boldsymbol{v}$, which changes with time from an initial state $\boldsymbol{p}^{(t_0)}$ to a final state $\boldsymbol{p}^{(t_1)}$. Player ${\cal X}$ (${\cal Y}$) controls probability flows of $e_1$, $e_3$ ($e_2$, $e_4$), which generate its partial entropy production $\Sigma_X$ ($\Sigma_Y$). In the final state, both the players take a penalty $D$ generated by an error between the final state $\boldsymbol{p}^{(t_1)}$ and the target state $\boldsymbol{p}^{\rm f}$ (black bars).
    }
    \label{F00}
\end{figure*}
\subsection{Markov jump process in bipartite model}
We consider a system consisting of two subsystems $X$ and $Y$, with binary states $x\in\{0,1\}$ and $y\in\{0,1\}$. A probability distribution of state $z=(x,y)\in\{0,1\}^2$ at time $t$ is denoted as $p_z^{(t)}$. In this system, a Markov jump process from $t=t_0$ to $t=t_1$ is described as a master equation;
\begin{align}
    \frac{d}{dt}p_z^{(t)}=&\sum_{z'}J_{z'\rightarrow z}^{(t)},
    \label{E01}\\
    J_{z'\rightarrow z}^{(t)}:=&W_{z'\rightarrow z}^{(t)}p_{z'}^{(t)}-W_{z\rightarrow z'}^{(t)}p_{z}^{(t)},
    \label{E02}
\end{align}
where $W_{z'\rightarrow z}^{(t)}$ and $J_{z'\rightarrow z}^{(t)}$ are a transition rate and a probability flow in a state transition from $z'=(x',y')\in\{0,1\}^2$ to $z$ at time $t$. By assuming a bipartite condition~\cite{strasberg2013thermodynamics,barato2013autonomous, horowitz2014thermodynamics,hartich2014stochastic, yamamoto2016linear}
that the state transition is independently performed in the subsystems $X$ and $Y$, the transition rate satisfies
\begin{subequations}
\begin{empheq}[left={W_{z'\rightarrow z}^{(t)}\empheqlbrace}]{alignat=2}
	=0 & \quad (x\neq x', y\neq y'),
	\label{E03a}\\
	\neq 0 & \quad ({\rm otherwise}).
	\label{E03b}
\end{empheq}
\end{subequations}
Because $W_{z'\rightarrow z}^{(t)}$ should not be zero if $W_{z\rightarrow z'}^{(t)} \neq 0$, the absolute irreversible transitions are not assumed.

To simplify the notation, we introduce a matrix representation as $(\boldsymbol{p}^{(t)})_i=p_{v_i}^{(t)}$ and the probability flow as $(\boldsymbol{J}^{(t)})_i=J_{e_i}^{(t)}$, where $\boldsymbol{v}:=((0,0),(1,0),(1,1),(0,1))$ represents a state $z=(x,y)$, and $(\boldsymbol{e})_i=v_{i}\to v_{\sigma(i)}$ represents a directed pair of states $(z\to z')$ which has a non-zero transition rate. Here, $\sigma(i)$ denotes a permutation of node, i.e., $\sigma(1)=2$, $\sigma(2)=3$, $\sigma(3)=4$, and $\sigma(4)=1$. Then, the master equation is rewritten as
\begin{align}
    \frac{d}{dt} \boldsymbol{p}^{(t)}&=\mathsf{B} \boldsymbol{J}^{(t)},
    \label{E04}\\
    \mathsf{B}_{ij}&= \delta_{i\sigma(j)} - \delta_{ij},
    \label{E05}
\end{align}
where $\delta_{ij}$ is the Kronecker delta and $\mathsf{B}$ is an incidence matrix. We remark that $\boldsymbol{v}$ implies a set of nodes, and $\boldsymbol{e}$ implies a set of directed edges in terms of Markov networks.

\subsection{Game-theretical formulation}
We further introduce players $\mathcal{X}$ and $\mathcal{Y}$ (see Fig. \ref{F01}-(a)). Player $\mathcal{X}$ can control the set of the transition rates $\hat{W}_X=\{W_{e}^{(t)}|t_0\le t\le t_1,e\in\mathcal{E}_X,e^{\dagger}\in\mathcal{E}_X\}$, that are relevant to transitions in subsystem $X$, i.e., $\mathcal{E}_X=\{e_1,e_3\}$. In the same way, player $\mathcal{Y}$ can control $\hat{W}_Y=\{W_{e}^{(t)}|t_0\le t\le t_1,e\in\mathcal{E}_Y,e^{\dagger}\in\mathcal{E}_Y\}$ with $\mathcal{E}_Y=\{e_2,e_4\}$ where $e^{\dagger}=z\to z'$ indicates a directed edge in the opposite direction of $e=z'\to z$.

This study proposes a problem of how the players cooperatively bring the final state $\boldsymbol{p}^{(t_1)}$ closer to the target $\boldsymbol{p}^{\rm f}$ from a given initial condition $\boldsymbol{p}^{(t_0)}=\boldsymbol{p}^{\rm i}$ as efficiently as possible (see Fig.~\ref{F00}). To deal with this problem, we now introduce two types of costs. The first cost is relevant to the ongoing processes of state transition in each subsystem. Players ${\cal X}$ and ${\cal Y}$ independently cost amounts of $\Sigma_X$ and $\Sigma_Y$, given by the partial entropy productions for a bipartite condition~\cite{horowitz2014thermodynamics,hartich2014stochastic}
\begin{align}
    \Sigma_X&:=\sum_{e_i\in\mathcal{E}_X}\int_{t_0}^{t_1} dt J_{e_i}^{(t)} F_{e_i}^{(t)},
    \label{E06}\\
    \Sigma_Y&:=\sum_{e_i\in\mathcal{E}_Y}\int_{t_0}^{t_1} dt J_{e_i}^{(t)} F_{e_i}^{(t)},
    \label{E07}\\
    F_{z'\to z}^{(t)}&:=\ln\frac{W_{z'\to z}^{(t)}p_{z'}^{(t)}}{W_{z\rightarrow z'}^{(t)}p_{z}^{(t)}}.
    \label{E08}
\end{align}
Here, $F_{z'\to z}^{(t)}$ is called a thermodynamic force. $\Sigma_X$ ($\Sigma_Y$) is always non-negative and gives $0$ if and only if $J_{e_i}^{(t)}=F_{e_i}^{(t)}=0$ for $e_i\in \mathcal{E}_X$ ($e_i\in \mathcal{E}_Y$.). We can define the total entropy production as the sum of the partial entropy productions $\Sigma_{\rm tot} =\Sigma_X + \Sigma_Y$. This total entropy production quantifies the dissipation in the total system during the time interval from $t=t_0$ to $t=t_1$~\cite{schnakenberg1976network, seifert2012stochastic} because the nonnegativity of the total entropy production $\Sigma_{\rm tot} \geq 0$ can be regarded as the second law of thermodynamics. The entropy production is widely discussed in terms of the stability of the system~\cite{schnakenberg1976network,prigogine1977self,mou1986stochastic,maes2015revisiting,ito2022information}, the fluctuation of the observable~\cite{barato2015thermodynamic, horowitz2020thermodynamic}, the speed of the time evolution~\cite{aurell2012refined,van2021geometrical,nakazato2021geometrical,yoshimura2023housekeeping}, the accuracy of the information transmission~\cite{allahverdyan2009thermodynamic,sagawa2012fluctuation,still2012thermodynamics,ito2013information,horowitz2014thermodynamics, hartich2014stochastic} and so on. The partial entropy productions $\Sigma_X$ and $\Sigma_Y$ can also be interpreted as the dissipation in the subsystems $X$ and $Y$, respectively.

The other cost is an error penalty of $D$ to both players due to failing to achieve the target state at the end. We introduce this penalty as the Pearson's chi-square divergence~\cite{pearson1900x}, which is used in a statistical test as a degree of current state $\boldsymbol{p}^{(t_1)}$ achieving the target $\boldsymbol{p}^{\rm f}$,
\begin{align}
    D&:=\frac{1}{2}\sum_{i}\frac{|p_{v_i}^{(t_1)}-p_{v_i}^{\rm f}|^2}{p_{v_i}^{\rm f}}.
    \label{E09}
\end{align}
Here, $D$ is always non-negative and gives $0$ if and only if the final and target states match, i.e., $\boldsymbol{p}^{(t_1)}=\boldsymbol{p}^{\rm f}$. We remark that this error penalty $D$ is a kind of the standard $f$-divergence~\cite{amari2016information}, which provides the Fisher information $2D= \sum_i (dp_{v_i}^{\rm f})^2/p_{v_i}^{\rm f}+O((dp_{v_i}^{\rm f})^3)$ for the small change $dp_{v_i}^{\rm f}=|p_{v_i}^{(t_1)}-p_{v_i}^{\rm f}|$. Thus, the result in this study is robust against the choice of the error penalty at least if the error penalty is given by the standard $f$-divergence and we only consider the small change $dp_{v_i}^{\rm f}$. For example, we can replace the error penalty $D$ with the Kullback-Leibler divergence $D_{\rm KL}=\sum_{i}p_{v_i}^{(t_1)}\ln(p_{v_i}^{(t_1)}/p_{v_i}^{\rm f}) \simeq D+O((dp_{v_i}^{\rm f})^3)$, which is a kind of the standard $f$-divergence widely used in thermodynamics~\cite{schlogl1971stability, schnakenberg1976network, jiu1984stability,esposito2010three,maes2015revisiting,ito2022information}.

For such costs $\Sigma_X$, $\Sigma_Y$, and $D$, we now consider a problem that each player $\mathcal{X}$ and $\mathcal{Y}$ aims to achieve the target state with as small partial entropy production as possible. Such a problem is given by two minimizations
\begin{subequations}
\begin{empheq}[left={\empheqlbrace}]{alignat=1}
	\min_{\hat{W}_{X}}\underbrace{\left[\Sigma_X+\lambda_XD\right]}_{=:C_X},
    \label{E10}\\
	\min_{\hat{W}_{Y}}\underbrace{\left[\Sigma_Y+\lambda_YD\right]}_{=:C_Y}.
    \label{E11}
\end{empheq}
\end{subequations}
Here, we introduce the penalty parameter, $\lambda_X$ or $\lambda_Y$, as the importance of achieving the target state for each subsystem. This study aims to find a Nash equilibrium solution~\cite{nash1950equilibrium} that satisfies minimization of costs $C_X$ and $C_Y$ simultaneously. We remark that $-C_X$ and $-C_Y$ correspond to the payoffs in game theory, and we consider the game that maximizes $-C_X$ and $-C_Y$ by changing the rate matrices $\hat{W}_{X}$ and $\hat{W}_{Y}$, and its optimal solution is obtained by a Nash equilibrium solution.

\section{Overview of main result}
We now overview the main results of this study. In order to obtain the Nash equilibrium solution, we assume a near-equilibrium condition, where (the reciprocal of) the Onsager coefficient $\alpha_{e_i}$ is defined for the transition $e_i$ ~\cite{schnakenberg1976network}. We remark that the words ``equilibrium'' and ``near-equilibrium condition'' without the word ``Nash'' are used only for thermal equilibrium that is introduced by the detailed balance condition. To avoid confusion, we also use the term ``Nash equilibrium'' every time without abbreviation for distinction. We also define summations for each subsystem as $\alpha_X=\sum_{e_i \in \mathcal{E}_X} \alpha_{e_i}$ and $\alpha_Y=\sum_{e_i \in \mathcal{E}_Y} \alpha_{e_i}$.

Our goal is to compute the partial entropy productions of the subsystem $X$ and $Y$ in the Nash equilibrium, i.e., $\Sigma_{X}^{\rm N}$ and $\Sigma_{Y}^{\rm N}$, especially when the given state transition is completely achieved $\boldsymbol{p}^{(t_1)}=\boldsymbol{p}^{\rm f}$
in the limit $\lambda_X\to\infty$ with the fixed ratio of importance $r:=\lambda_X/\lambda_Y$, or equivalently in the limit $\lambda_Y\to\infty$ with the fixed ratio $r$. We obtained the main results as follows,
\begin{align}
    \Sigma_{X}^{\rm N}&=\Sigma_{X}^{\rm min}+f_X(r; \gamma)\Sigma_{XY} \ge \Sigma_{X}^{\rm min},
    \label{E12}\\
    \Sigma_{X}^{\rm N}&= \Sigma_{X}^{\rm min}\hspace{1cm} (r \to 0),
    \label{E13}\\
    \Sigma_{Y}^{\rm N}&=\Sigma_{Y}^{\rm min}+f_Y(r; \gamma)\Sigma_{XY}\ge \Sigma_{Y}^{\rm min},
    \label{E14}\\
    \Sigma_{Y}^{\rm N}&=\Sigma_{Y}^{\rm min}\hspace{1cm} (r \to \infty),
    \label{E15}
\end{align}
where $\Sigma_{X}^{\rm min}$ ($\Sigma_{Y}^{\rm min}$) is the minimum partial entropy productions of $X$ ($Y$), and $\Sigma_{XY}$ is the shared minimum entropy production, which is given by the minimum total entropy production
\begin{align}
    \Sigma_{\rm tot}^{\rm N} &:= \Sigma_{X}^{\rm N}+\Sigma_{Y}^{\rm N}\ge\Sigma_{X}^{\rm min}+\Sigma_{Y}^{\rm min}+\Sigma_{XY} =: \Sigma_{\rm tot}^{\rm min}.
    \label{E16}\\
     \Sigma_{\rm tot}^{\rm N}&=\Sigma_{\rm tot}^{\rm min}\hspace{1cm} (r = 1),
    \label{E16_2}
    \end{align}
The factor $f_X(r; \gamma)(\ge 0)$ ($f_Y(r; \gamma)(\ge 0)$) that satisfies $f_X(0; \gamma)= f_Y(\infty; \gamma) =0$ is the function of $\gamma:=\alpha_X/\alpha_Y$ which is monotonically increasing (decreasing) with $r$. Thus, the minimum partial entropy production $\Sigma_{X}^{\rm min}$ ($\Sigma_{Y}^{\rm min}$) is achieved when $r\to 0$ ($r\to\infty$). $f_X(r; \gamma)$ and $f_Y(r; \gamma)$ are also functions of $\gamma:=\alpha_X/\alpha_Y$. For $r=1$, the factor is given by $f_X(1; \gamma)=1/(\gamma+1)$, $f_Y(1; \gamma)=\gamma/(\gamma+1)$, and the minimum total entropy production is achieved $\Sigma_{\rm tot}^{\rm N}= \Sigma_{\rm tot}^{\rm min}$ when $r=1$.

The above result implies that there exists a trade-off relation between the partial entropy productions such that a partial entropy production should be larger if we want the other partial entropy production to be smaller. Moreover, the total entropy production can be minimized if each subsystem equally shares the penalty. That implies that an equivalent penalty in two subsystems reduces the dissipation in the total system, and the minimization of dissipation in only one subsystem increases the dissipation of the total system.

\section{Physical importance and example}
\subsection{Physical interpretations based on information-energy conversion}
Our results on the lower bound of the partial entropy productions can be immediately applied to the topic of the efficiency of the information-energy conversion in terms of Maxwell's demon. To discuss the information-energy conversion, we first decompose the thermodynamics force, i.e., Eq.~\eqref{E08}, into the thermodynamic part ($\rm T$) and information part ($\rm I$) as
\begin{eqnarray}
F_{z'\to z}^{(t)} &=& F_{z'\to z}^{{\rm T}(t)} +  F_{z'\to z}^{{\rm I}(t)}, \\
F_{z'\to z}^{{\rm T}(t)} &=& \ln \frac{W_{z'\to z}^{(t)}}{W_{z\rightarrow z'}^{(t)}}, \\
F_{z'\to z}^{{\rm I}(t)} &=& \ln p_{z'}^{(t)} - \ln p_{z}^{(t)}.
\end{eqnarray}
The partial entropy production is also decomposed into two contributions~\cite{horowitz2014thermodynamics, yamamoto2016linear} 
\begin{eqnarray}
\Sigma_X = \Sigma_X^{\rm T} + \Sigma_X^{\rm I},
\end{eqnarray}
where the thermodynamic contribution 
\begin{eqnarray}
\Sigma_X^{\rm T} = \sum_{e_i\in\mathcal{E}_X}\int_{t_0}^{t_1} dt J_{e_i}^{(t)} F_{e_i}^{{\rm T}(t)},
\end{eqnarray}
means the entropy change of the heat bath, while the informational contribution
\begin{eqnarray}
\Sigma_X^{\rm I} = \sum_{e_i\in\mathcal{E}_X}\int_{t_0}^{t_1} dt J_{e_i}^{(t)} F_{e_i}^{{\rm I}(t)},
\end{eqnarray}
means the sum of the entropy change of the system and information flow because $F_{e_i}^{{\rm I}(t)}$ is given by the stochastic Shannon entropy change. For a bipartite system, the sum $\Sigma_X^{\rm I}+\Sigma_Y^{\rm I}$ is regarded as the change of the Shannon entropy in the total system $(X, Y)$, which is given by the sum of the Shannon entropy in $X$ and $Y$, and the mutual information between two subsystems $X$ and $Y$.

Because the partial entropy production $\Sigma_X \geq 0$ is non-negative, we obtain the inequality
\begin{eqnarray}
\Sigma_X^{\rm T} \geq - \Sigma_X^{\rm I},
\end{eqnarray}
which is called the second law of information thermodynamics. This result can explain the information-energy conversion by Maxwell's demon because this inequality explains a trade-off between the thermodynamic contribution  $\Sigma_X^{\rm T}$ and the informational contribution $- \Sigma_X^{\rm I}$, and the thermodynamic contribution $\Sigma_X^{\rm T}$ can be negative due to the information contribution $- \Sigma_X^{\rm I}$. The value of the partial entropy production $\Sigma_X$ is regarded as the dissipation in the information-energy conversion. For example, the equality $\Sigma_X^{\rm T} = - \Sigma_X^{\rm I}$ (or $\Sigma_X = 0$) holds for the Szilard engine, which is an optimal heat engine driven by Maxwell's demon to achieve the maximal efficiency for information-energy conversion~\cite{toyabe2010experimental}.

Thus, a minimization problem of $\Sigma_X$ can be interpreted as a maximization problem of the efficiency for information-energy conversion, and our study can be regarded as tackling a problem to maximize the efficiency for information-heat conversion under a restriction that the subsystems cooperatively achieve a given task within a finite time. 
Indeed, we obtained a positive bound for the partial entropy production 
\begin{eqnarray}
 \Sigma_{X}\ge\Sigma_{X}^{\rm min} \Leftrightarrow \Sigma_X^{\rm T}\ge \Sigma_{X}^{\rm min}-\Sigma_X^{\rm I},
\end{eqnarray}
which means that there is an inevitable dissipation $\Sigma_{X}^{\rm min}$ in the information-energy conversion caused by the finite-time task.

\subsection{Example: {\it E. coli} chemotaxis and game-theoretic interpretation of {\it E. coli} evolution}
\begin{figure*}
    \centering
    \includegraphics[width=0.7\hsize]{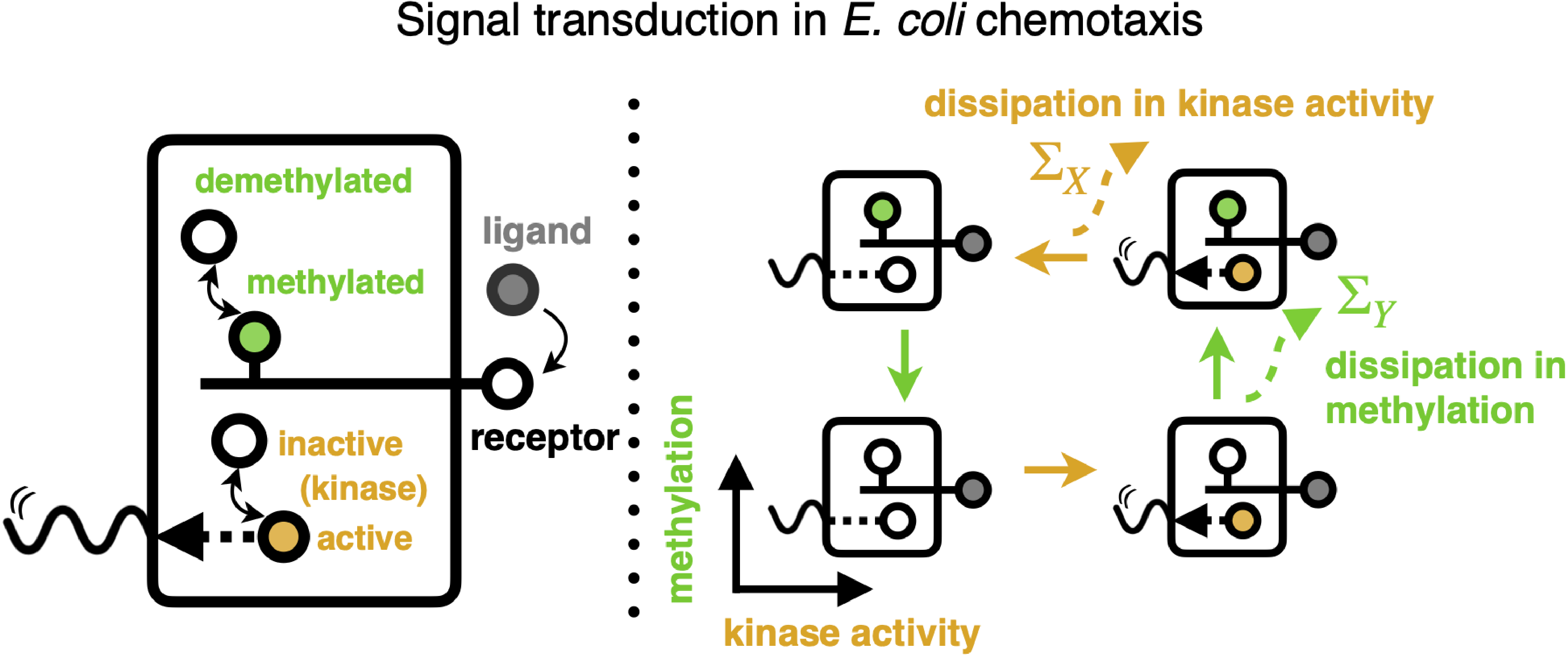}
    \caption{A physical example of a bipartite system, signal transduction of {\it E. coli} chemotaxis. Subsystem $X$ takes an active (orange-colored dot) or inactive (uncolored dot) state of a kinase. The other subsystem $Y$ takes a methylated (green-colored dot) or demethylated (uncolored dot) state of the receptor. During a state transition, heat dissipation emerges for each subsystem as the partial entropy production.
    }
    \label{F01}
\end{figure*}

This study might be applied to how {\it E. coli} evolves in its adaptation process in chemotaxis. To discuss the applicability, we first explain {\it E. coli} chemotaxis based on the bipartite model~\cite{tu2008nonequilibrium} (see Fig.~\ref{F01} for visualization). The bipartite model of {\it E. coli} chemotaxis is organized by the activity of kinase (CheA), the methylation level of the receptor, and the ligand concentration in the environment. Let subsystem $X$ denote whether the kinase is inactive ($x=0$) or active ($x=1$). On the other hand, let subsystem $Y$ denote whether the receptor is demethylated ($y=0$) methylated ($y=1$). The activation of the kinase and the methylation of the receptor independently occurred stochastically. Thus, Eqs.~\eqref{E03a}-\eqref{E03b} is satisfied, and the master equation is given by Eqs.~\eqref{E01}-\eqref{E02}. In stochastic thermodynamics, heat dissipation can be introduced based on the following local detailed balance conditions, 
\begin{eqnarray}
    \ln \frac{W_{(x',y)\to (x,y)}^{(t)}}{W_{(x,y)\to (x',y)}^{(t)}} = -\beta_X \Delta Q_X^{(x',y)\to (x,y)}, \\
    \ln \frac{W_{(x,y')\to (x,y)}^{(t)}}{W_{(x,y)\to (x,y')}^{(t)}} = -\beta_Y \Delta Q_Y^{(x,y')\to (x,y)},
\end{eqnarray}
where $\beta_X$ and $\beta_Y$ are inverse temperatures in the kinase and the receptor, $\Delta Q_X^{(x',y)\to (x,y)}$ is the heat dissipation of the kinase from state $x'$ to state $x$ under the condition of $y$, and $\Delta Q_Y^{(x,y')\to (x,y)}$ is the heat dissipation of the receptor from state $y'$ to state $y$ under the condition of $x$. The heat dissipation $\Delta Q_X^{(x',y)\to (x,y)}$ can be given by the difference between the energy and chemical potential~\cite{yamamoto2016linear}. Because the activation of the kinase is driven by the receptor-ligand binding, the heat dissipation $\Delta Q_X^{(x',y)\to (x,y)}$ generally depends on the ligand concentration in the environment.

We here explain typical dynamics of adaptation in {\it E. coli} chemotaxis with this 4-state model. The adaptation is transient dynamics after the ligand concentration is changed. At time $t=t_0$, we assume that the initial state stochastically tends to be $(1,0)$. This means that the kinase tends to be active, and the receptor tends to be demethylated. We here assume that the ligand concentration is changed from $L=0$ to $L=1$ at time $t=t_0$, where $L=0$ means that the ligand concentration is relatively low, and $L=1$ means that the ligand concentration is relatively high, respectively. At time $t>t_0$, the kinase activity can change from active state $x=1$ to inactive state $x=0$ rapidly, and this change activates other kinases CheY. The activity of CheY changes the mode of the flagellar motor from the tumbling mode to the run mode, and this mode change in the flagellar motor explains the behavior of chemotaxis. From time $t=t_0$ to $t=t_1$, the kinase activity gradually returns to the active state $x=1$, and the mode of the flagellar motor returns to the tumbling mode. This behavior is regarded as the adaption in {\it E. coli} chemotaxis because {\it E. coli} adapts to the environmental change from $L=0$ to $L=1$. In summary, the state stochastically tends to change in this adaptation dynamics as
\begin{eqnarray}
(1,0) \to (0,0) \to (0, 1) \to (1, 1).
\end{eqnarray}
In terms of probability distribution, the adaptation dynamics is given by the initial condition $p_{v_i}^{(t_0)}$ and final condition $p_{v_i}^{(t_1)}$ such that $p_{v_2}^{(t_0)}$ is relatively large rather than $p_{v_1}^{(t_0)}$, $p_{v_3}^{(t_0)}$ and $p_{v_4}^{(t_0)}$, and  $p_{v_3}^{(t_1)}$ is relatively large rather than $p_{v_1}^{(t_1)}$, $p_{v_2}^{(t_1)}$ and $p_{v_4}^{(t_1)}$, respectively. 

We now consider evolutionary processes to minimize dissipation in the kinase and the receptor under the constraint of precise adaptation. Because the partial entropy production, especially the thermodynamic contributions $\Sigma_X^{\rm T}$ and $\Sigma_Y^{\rm T}$, are related to the heat dissipation
\begin{eqnarray}
    \Sigma_X^{\rm T} = -\sum_{e_i\in\mathcal{E}_X}\int_{t_0}^{t_1} dt J_{e_i}^{(t)} \beta_X \Delta Q_X^{e_i}, \\
    \Sigma_Y^{\rm T} = -\sum_{e_i\in\mathcal{E}_Y}\int_{t_0}^{t_1} dt J_{e_i}^{(t)} \beta_Y \Delta Q_Y^{e_i},
\end{eqnarray}
and $\Sigma_X=\Sigma_X^{\rm T}+ \Sigma_X^{\rm I} $ ($\Sigma_Y=\Sigma_Y^{\rm T}+ \Sigma_Y^{\rm I}$) can be interpreted as the local free energy difference under the condition of the other system $Y$ ($X$)~\cite{sagawa2010generalized,sagawa2012fluctuation}, the minimization of the partial entropy production is related to the savings of the local free energy as a thermodynamic resource. For {\it E. coli}, the thermodynamic resource like feed can be obtained as a consequence of chemotaxis. Thus, {\it E. coli} is motivated to achieve the task of chemotaxis precisely to gain the thermodynamic resource. This achievement of the task is measured by an error penalty $D$ given by Eq.~\eqref{E09} for adaptation dynamics from $t=t_0$ to $t=t_1$. To gain feed as a thermodynamic resource from the environment, {\it E. coli} would like to minimize this error penalty $D$, and its thermodynamic contributions of the penalty for the kinase ($X$) and the receptor ($Y$) can be treated independently by multiplying the importance of achieving the target state in the adaptation dynamics such as $\lambda_X D$ and $\lambda_Y D$. Thus, $C_X=\left[\Sigma_X+\lambda_XD\right]$ and $C_Y=\left[\Sigma_Y+\lambda_YD\right]$ correspond to the net thermodynamic dissipations for the kinase and the receptor, respectively. The minimization problems of $C_X$ and $C_Y$ are needed for {\it E. coli} to efficiently survive an environment with low feed. In evolutionary processes, the phenotype of the kinase and the receptor can be changed accidentally, and the difference of the phenotype in adaptation dynamics can be quantified as the set of the transition rates $\hat{W}_X$ for the kinase and $\hat{W}_Y$ for the receptor. In evolutionary processes, the phenotype can be changed to maximize the fitness of {\it E. coli}. To survive an environment with low feed, the fitness can be $-C_X$ and $-C_Y$ as the net free energy gain. Because mutations of the phenotype for the kinase and the receptor can happen independently, we can consider two minimization problems, which are equivalent to the maximization problems of the fitness
\begin{eqnarray}
&{\rm min}_{\hat{W}_X} C_X =-  {\rm max}_{\hat{W}_X}(-C_X) \label{minimum1sup}, \\
&{\rm min}_{\hat{W}_X} C_Y =-  {\rm max}_{\hat{W}_Y}(-C_Y)  \label{minimum2sup}.
\end{eqnarray}
If we assume that the current phenotype of the receptor and the kinase $\hat{W}_X$ and $\hat{W}_Y$ in adaptation dynamics is obtained as a consequence of evolutionary processes, we can assume that the dissipation $\Sigma_X$ and $\Sigma_Y$ in the kinase and the methylation may be discussed in terms of the Nash equilibrium solutions $\Sigma_X^{\rm N}$ and $\Sigma_Y^{\rm N}$ for two minimization problems Eqs.~(\ref{minimum1sup}) and (\ref{minimum2sup}). 

Based on the above evolutionary processes, we can discuss a possible validity of our game-theoretic framework in {\it E. coli} chemotaxis. Now, we may assume that the transition from the initial state to the final state for the methylation level of the receptor is more important than that for the kinase activity in the adaptation dynamics. That is because the methylation level tends to be changed from the demethylated state $y=0$ at time $t=t_0$ to the methylated state $y=1$ at time $t=t_1$ during the adaptation dynamics. On the other hand, the kinase activity is only instantaneously changed during the adaptation dynamics, and the initial state $x=1$ at time $t=t_0$ tends to be the same as the final state $x=1$ at time $t=t_1$. Thus, achievement of the target state $p_{v_i}^{(t_1)}$ may be more important for the receptor than for the kinase in adaptation dynamics. In our game-theoretic framework, the difference in the importance may be quantified as $\lambda_Y \gg \lambda_X$. 
Experimentally, $\Sigma_Y \gg \Sigma_X$ can be seen in adaptation dynamics of {\it E. coli} because the relaxation time of the receptor is relatively slower than the relaxation time of the kinase~\cite{tu2008nonequilibrium, ito2015maxwell}. This fact is consistent with our main result that $\Sigma^{\rm N}_Y \gg \Sigma^{\rm N}_X$ for $\lambda_Y \gg \lambda_X$. Thus, our main result may provide a possible explanation of the huge difference in thermodynamic dissipation of each subsystem in the signal transduction of {\it E. coli} chemotaxis, which is based on the game-theoretic evolution in an environment with low feed under the constraint of the precise adaptation dynamics. 

\section{Nash equilibrium solution in linear irreversible thermodynamics}
In general, the problem of Eqs.~(\ref{E10}) and (\ref{E11}) does not give a non-trivial conclusion without any constraint because the entropy production can be zero without any constraint for the Markov jump process. In this study, we consider a constraint of the near-equilibrium condition in linear irreversible thermodynamics~\cite{schnakenberg1976network,prigogine1977self}. In linear irreversible thermodynamics, we assume that the transition rate is given by $W_{z'\rightarrow z}^{(t)}=W_{z'\rightarrow z}^{\rm eq}+O(\delta W)$ where $\delta W$ is the small change of the transition rate and $W_{z'\rightarrow z}^{\rm eq}$ satisfies the detailed balance condition $W_{z'\rightarrow z}^{\rm eq}p_{z'}^{\rm eq}=W_{z\rightarrow z'}^{\rm eq}p_{z}^{\rm eq}$ with the equilibrium distribution $p_{z}^{\rm eq}$ for all pairs of $z$ and $z'$. We also assume that the initial state satisfies the near-equilibrium condition. For example, $\boldsymbol{p}^{(t_0)}$ is the steady state distribution for $W_{z'\rightarrow z}^{\rm eq}+O(\delta W)$. In the above setup, we can confirm $\boldsymbol{J}^{(t)}=O(\delta W)$ and $\boldsymbol{p}^{(t)}=\boldsymbol{p}^{\rm eq}+O(\delta W)$ during the transition process $t_0 \leq t \leq t_1:= t_0+ \tau$.

We here explain the Nash equilibrium solution (see also Appendix~A-C for detailed derivation). Let $\alpha_{z'\rightarrow z}:=(W_{z'\rightarrow z}^{\rm eq}p_{z'}^{\rm eq})^{-1}=(W_{z\rightarrow z'}^{\rm eq}p_{z}^{\rm eq})^{-1}$ be (the reciprocal of) the Onsager coefficient. $\Sigma_X$ and $\Sigma_Y$ become quadratic functions by ignoring terms $O(\delta W^3)$, and thus the lower bounds are given by $\Sigma_X \geq \tau \sum_{e_i \in \mathcal{E}_X} \alpha_{e_i} (\bar{J}_{e_i}^{(t)})^2$ and $\Sigma_Y \geq \tau \sum_{e_i \in \mathcal{E}_Y} \alpha_{e_i} (\bar{J}_{e_i}^{(t)})^2$ where $\bar{J}_{e_i}$ is time-averaged flow defined as $\bar{J}_{e_i}:=(\int_{t_0}^{t_0+\tau}dt J_{e_i}^{(t)})/\tau$. The penalty $D$ is also given by the quadratic function of $\bar{J}_{e_i}$. Thus, we can describe the lower bounds of cost $\bar{C}_X(\bar{J}_X, \bar{J}_Y) (\leq C_X)$ and $\bar{C}_Y(\bar{J}_X, \bar{J}_Y) (\leq C_Y)$ as functions of time-averaged flows $\bar{J}_X:=\{\bar{J}_{e_i}|e_i\in\mathcal{E}_X\}$ and $\bar{J}_Y:=\{\bar{J}_{e_i}|e_i\in\mathcal{E}_X\}$. Optimal time-averaged flows to minimize the cost are described by
\begin{subequations}
\begin{empheq}[left={\empheqlbrace}]{alignat=1}
	\bar{J}_X^*(\bar{J}_Y)=&{\rm arg} \min_{\bar{J}_{X}} \bar{C}_X(\bar{J}_X,\bar{J}_Y),
    \label{E28}\\
	\bar{J}_Y^*(\bar{J}_X)=&{\rm arg} \min_{\bar{J}_{Y}} \bar{C}_Y(\bar{J}_X,\bar{J}_Y).
    \label{E29}
\end{empheq}
\end{subequations}
Thus, the Nash equilibrium solution for time-averaged flows $(\bar{J}_X^{\rm N},\bar{J}_Y^{\rm N})=(\{\bar{J}_{e_1}^{\rm N},\bar{J}_{e_3}^{\rm N}\},\{\bar{J}_{e_2}^{\rm N},\bar{J}_{e_4}^{\rm N}\})$ satisfies
\begin{subequations}
\begin{empheq}[left={\empheqlbrace}]{alignat=1}
	\bar{J}_X^{\rm N}=\bar{J}_X^{*}(\bar{J}_Y^{\rm N}),
    \label{E26}\\
	\bar{J}_Y^{\rm N}=\bar{J}_Y^{*}(\bar{J}_X^{\rm N}),
    \label{E27}
\end{empheq}
\end{subequations}
as the fixed point of Eqs.~(\ref{E28}) and (\ref{E29}).

The analytical calculation of the Nash equilibrium solution is as follows (see also Appendix~A-C). From Eqs.~(\ref{E26})-(\ref{E29}), $\bar{J}_X^{\rm N}$ and $\bar{J}_Y^{\rm N}$ satisfy extreme value conditions for $C_X(\bar{J}_X,\bar{J}_Y)$ and $C_Y(\bar{J}_X,\bar{J}_Y)$, respectively. Because $C_X(\bar{J}_X,\bar{J}_Y)$ and $C_Y(\bar{J}_X,\bar{J}_Y)$ are quadratic functions of $(\bar{\boldsymbol{J}}^{\rm N})_i:=\bar{J}_{e_i}^{\rm N}$, the solution of the extreme value conditions $\bar{\boldsymbol{J}}^{\rm N}$ are analytically obtained. Especially in the limit $\lambda_X{\to}\infty$ with the fixed ratio $r$, $\bar{\boldsymbol{J}}^{\rm N}$ are given by
\begin{align}
   \bar{J}_{e_i}^{\rm N} =& \frac{1}{{\tau(\alpha_X + r \alpha_Y})} \left[ -r \alpha_{e_{\sigma^3(i)}}\delta p_{v_{i}} \right. \nonumber  \\
   &\left. +(r \alpha_{e_{\sigma(i)}}+\alpha_{e_{\sigma^2(i)}})\delta p_{v_{\sigma(i)}} +\alpha_{e_{\sigma^2(i)}}\delta p_{v_{\sigma^2(i)}} \right],
   \label{E34}
\end{align}
for $e_i\in{\cal E}_X$, and
\begin{align}
   \bar{J}_{e_i}^{\rm N} =& \frac{1}{{\tau (\alpha_X + r \alpha_Y})} \left[ -\alpha_{e_{\sigma^3(i)}}\delta p_{v_{i}} \right. \nonumber  \\
   &\left. +(\alpha_{e_{\sigma(i)}}+r\alpha_{e_{\sigma^2(i)}})\delta p_{v_{\sigma(i)}} +r\alpha_{e_{\sigma^2(i)}}\delta p_{v_{\sigma^2(i)}} \right],
   \label{E35}
\end{align}
for $e_i\in{\cal E}_Y$, where the permutation $\sigma^2(i)$ and $\sigma^3(i)$ are defined as $\sigma^2(i)=\sigma(\sigma(i))$ and $\sigma^3 (i)=\sigma(\sigma(\sigma(i)))$, respectively.

\section{Minimum partial entropy productions for the Nash equilibrium solution}
We discuss the minimum partial entropy productions for the Nash equilibrium solution, that is $(\Sigma_X^{\rm N},\Sigma_Y^{\rm N}):=(\Sigma_X,\Sigma_Y)|_{\boldsymbol{J}^{(t)}=\bar{\boldsymbol{J}}^{\rm N}}$. In the limit $\lambda_X{\to}\infty$ with the fixed ratio $r$, $\Sigma_X^{\rm N}=\Sigma_{X}^{\rm min}+f_X(r; \gamma)\Sigma_{XY}$ and $\Sigma_Y^{\rm N}=\Sigma_{Y}^{\rm min}+f_Y(r; \gamma)\Sigma_{XY}$ in Eqs. (\ref{E12}) and (\ref{E14}) are analytically given by
\begin{align}
    &\Sigma_X^{\rm min}=\frac{(\int_{t_0}^{t_0+\tau} dt \bar{\mathcal{J}}_{X}^{\rm N})^2}{\tau(\sum_{e_i\in{\cal E}_X}\alpha_{e_i}^{-1} )},\ \Sigma_Y^{\rm min}=\frac{(\int_{t_0}^{t_0+\tau} dt \bar{\mathcal{J}}_{Y}^{\rm N})^2}{\tau(\sum_{e_i\in{\cal E}_Y}\alpha_{e_i}^{-1})},
    \label{E36}\\
    &\Sigma_{XY}=\frac{(\int_{t_0}^{t_0+\tau} dt \bar{\mathcal{J}}_{XY}^{\rm N})^2}{\tau(\alpha_X^{-1}+\alpha_Y^{-1})},
    \label{E37}\\
    &f_X(r;\gamma)=\frac{r^2(\gamma+1)}{(\gamma+r)^2},\ f_Y(r;\gamma)=\frac{\gamma(\gamma+1)}{(\gamma+r)^2},
    \label{E38}
\end{align}
where the flow $\bar{\mathcal{J}}_{X}^{\rm N}$, $\bar{\mathcal{J}}_{Y}^{\rm N}$ and $\bar{\mathcal{J}}_{XY}^{\rm N}$ are the linear transformation of $\bar{\boldsymbol{J}}^{\rm N}$ defined as
\begin{align}
    \bar{\boldsymbol{\mathcal{J}}}^{\rm N}=&{\sf T} \bar{\boldsymbol{J}}^{\rm N},
    \label{E39}\\
    \bar{\boldsymbol{\mathcal{J}}}^{\rm N}:=&(\bar{\mathcal{J}}_{X}^{\rm N},\bar{\mathcal{J}}_{Y}^{\rm N},\bar{\mathcal{J}}_{XY}^{\rm N},\bar{\mathcal{J}}_{\rm rot}^{\rm N})^{\rm T},
    \label{E40}\\
    {\sf T} :=&\left(\begin{array}{cccc}
        1 & 0 & -1 & 0 \\
        0 & 1 & 0 & -1 \\
        \frac{\alpha_{e_1}}{\alpha_X} & -\frac{\alpha_{e_2}}{\alpha_Y} & \frac{\alpha_{e_3}}{\alpha_X} & -\frac{\alpha_{e_4}}{\alpha_Y} \\
        \frac{1}{4} & \frac{1}{4} & \frac{1}{4} & \frac{1}{4} \\
    \end{array}\right).
    \label{E41}
\end{align}
Here, $\bar{\mathcal{J}}_{X}^{\rm N}$ ($\bar{\mathcal{J}}_{Y}^{\rm N}$) implies the probability flow for the marginal distribution of $X$ ($Y$). $\bar{\mathcal{J}}_{XY}^{\rm N}$ represents the probability flow of the interaction between $X$ and $Y$, which cannot be written by $\bar{\mathcal{J}}_{X}^{\rm N}$ and $\bar{\mathcal{J}}_{Y}^{\rm N}$. The probability flow $\bar{\mathcal{J}}_{\rm rot}^{\rm N}$ does not contribute to the time evolution $d\boldsymbol{p}^{(t)}/dt = \mathsf{B} {\sf T}^{-1} \bar{\boldsymbol{\mathcal{J}}}^{\rm N}$ because $\mathsf{B} {\sf T}^{-1} (0,0,0, \bar{\mathcal{J}}_{\rm rot}^{\rm N})^{\rm T} = \boldsymbol{0}$, and thus $\bar{\mathcal{J}}_{\rm rot}^{\rm N}$ can be interpreted in terms of optimal transport theory~\cite{maas2011gradient}, which explains the minimization of the entropy production in a finite time~\cite{yoshimura2023housekeeping} (see Appendix~D). We remark that $\Sigma_X^{\rm min}$, $\Sigma_Y^{\rm min}$, $\Sigma_{XY}$ are proportional to $1/\tau$, and this fact is consistent with the thermodynamic speed limit based on optimal transport theory~\cite{aurell2012refined, dechant2019thermodynamic,van2021geometrical,nakazato2021geometrical,yoshimura2023housekeeping}. Indeed, Eq.~(\ref{E36}) can be interpreted as a thermodynamic speed for the partial entropy production (see Appendix~D). Moreover, there is one-to-one correspondence between $\bar{\mathcal{J}}_{\rm rot}^{\rm N}$ and $r$ (see Appendix~B). Thus, the minimization of the partial entropy production $\Sigma^N_X = \Sigma_X^{\rm min}$ ($\Sigma^N_Y = \Sigma_Y^{\rm min}$) and the minimization of the total entropy production $\Sigma^N_{\rm tot}= \Sigma_{\rm tot}^{\rm min}$ are achievable by changing $\bar{\mathcal{J}}_{\rm rot}^{\rm N}$ because the factors $f_X(r;\gamma)$ and $f_Y(r;\gamma)$ are determined by the flow $\bar{\mathcal{J}}_{\rm rot}^{\rm N}$.
\begin{figure}[htbp]
\begin{center}
\includegraphics[width=1.0\linewidth]{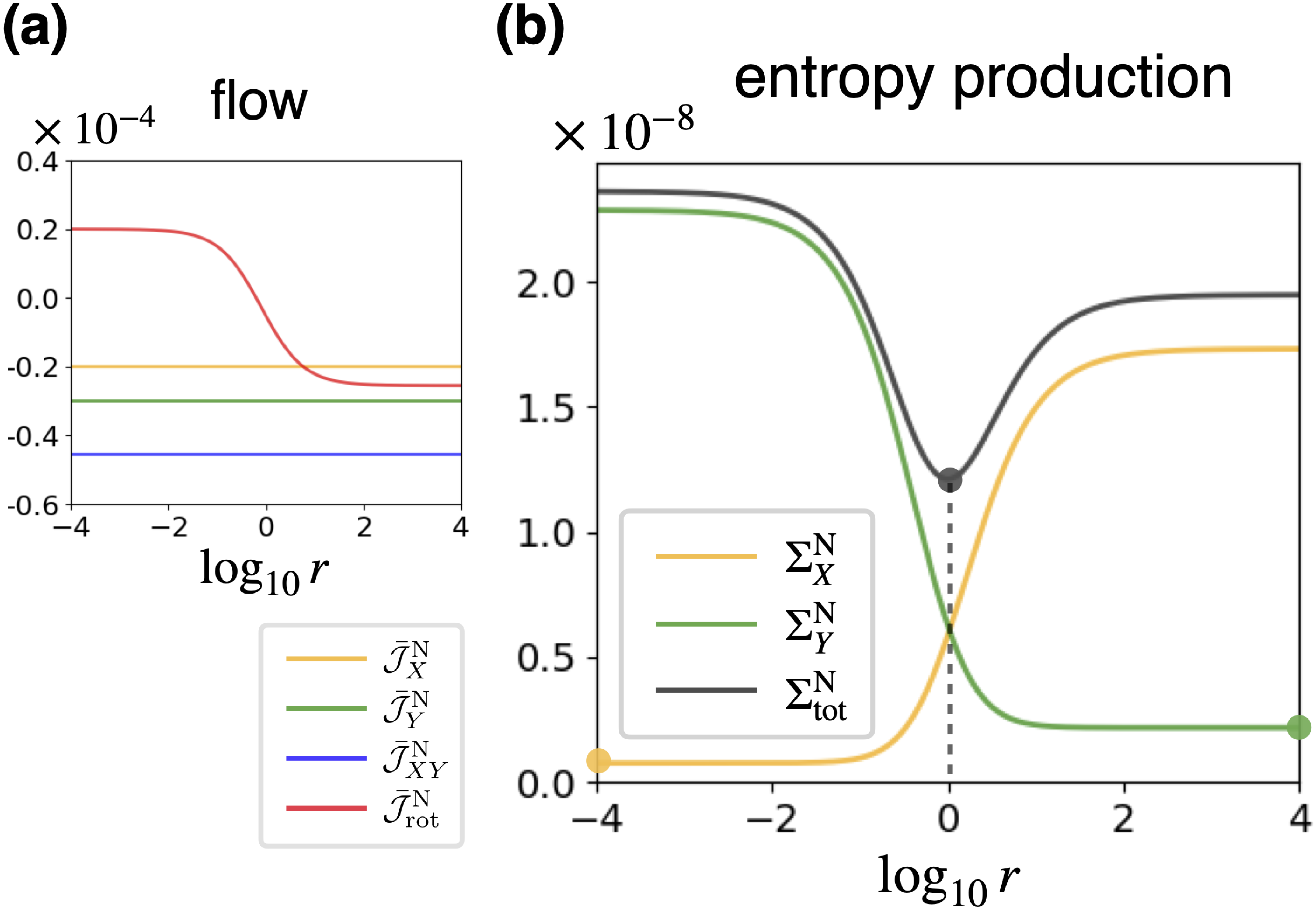}
\caption{(a) Probability flows and (b) partial entropy productions under $\lambda_X=\lambda_Y=\infty$. We give the environmental parameters as $(\alpha_{e_1},\alpha_{e_2},\alpha_{e_3},\alpha_{e_4})=(3,4,5,6)$, $\boldsymbol{p}^{\rm f}=(0.1,0.2,0.3,0.4)$, $\delta\boldsymbol{p}=(0.7,-0.4,0.2,-0.5)\times10^{-4}$, $(t_0,t_1)=(0,1)$. The horizontal axis is $\log_{10} r$. In panel (a), orange, green, blue, and red lines indicate $\bar{\mathcal{J}}_{X}^{\rm N}$, $\bar{\mathcal{J}}_{Y}^{\rm N}$, $\bar{\mathcal{J}}_{XY}^{\rm N}$, $\bar{\mathcal{J}}_{\rm rot}^{\rm N}$, respectively. In panel (b), orange, green, and black lines indicate $\Sigma_X^{\rm N}$, $\Sigma_Y^{\rm N}$, and $\Sigma_{\rm tot}^{\rm N}$, respectively. Then, each color dot indicates a minimum value of the cost.}
\label{F02}
\end{center}
\end{figure}

In Fig.~\ref{F02}, we illustrate a trade-off relation between $\Sigma_X^{\rm N}$ and $\Sigma_Y^{\rm N}$ by a behavior of probability flows $\bar{\mathcal{J}}_{X}^{\rm N}$ and the minimum entropy productions $\Sigma^{\rm N}_X$, $\Sigma^{\rm N}_Y$, $\Sigma^{\rm N}_{\rm tot}$ for the Nash equilibrium solution in the case of $\lambda_X{\to}\infty$ with the fixed $r:=\lambda_X/\lambda_Y$. As seen in Fig.~\ref{F02}(a), only $\bar{\mathcal{J}}^{\rm N}_{\rm rot}$ monotonically changes with $r$, and $\bar{\mathcal{J}}^{\rm N}_{X}$, $\bar{\mathcal{J}}^{\rm N}_{Y}$ and $\bar{\mathcal{J}}^{\rm N}_{XY}$ do not depend on $r$ (see also Appendix~B). As seen in Fig.~\ref{F02}(b), there is the trade-off relation between $\Sigma_X^{\rm N}$ and $\Sigma_Y^{\rm N}$ in terms of $r$. We can see that $\Sigma_{\rm tot}^N = \Sigma_{\rm tot}^{\rm min}$ when $r=1$, $\Sigma_{X}^N = \Sigma_{X}^{\rm min}$ when $r{\to}0$, and $\Sigma_{Y}^N = \Sigma_{Y}^{\rm min}$ when $r{\to}\infty$.

We briefly summarize results for cases of finite values of $\lambda_X$ and $\lambda_Y$ discussed in Appendix~E. For finite $\lambda_X$ and $\lambda_Y$, $\bar{\mathcal{J}}_{X}^{\rm N}$, $\bar{\mathcal{J}}_{Y}^{\rm N}$, $\bar{\mathcal{J}}_{XY}^{\rm N}$ generally depend on $r$. Nevertheless, we can see a trade-off relation between $\Sigma_X^{\rm N}$ and $\Sigma_Y^{\rm N}$ for finite $\lambda_X$ and $\lambda_Y$ because the dependence of $\bar{\mathcal{J}}_{\rm rot}^{\rm N}$, $\Sigma_{X}^{\rm N}$, and $\Sigma_{Y}^{\rm N}$ on $r$ is similar to the case of $\lambda_{X}{\to}\infty$ with the fixed ratio $r$.

\section{Conclusion and discussion}
This study incorporated a game-theoretic approach into the minimization problems of the partial entropy productions. As a representative example, we consider bipartite systems where two minimization problems of the partial entropy productions can be conflicted. We consider a problem that the partial entropy productions should be minimized for each subsystem with a given state transition for whole systems. We formulated the solution to this problem as Nash equilibrium and analytically calculated it. We found that both subsystems inevitably take their necessary partial entropy productions for achieving the task in a finite time. Lower bounds of the partial entropy productions for each subsystem and the total system are analytically derived. Interestingly, there is a trade-off relation between the minimum partial entropy productions of the subsystems. Furthermore, the minimum values of these partial entropy productions and the total entropy production cannot be achieved at the same time.

Our findings may apply to several interesting topics in thermodynamics. For example, our result may explain the minimum thermodynamic cost and optimal protocol in the finite-time information-energy conversion. The analytical expressions of the minimum entropy productions Eqs.~(\ref{E36}) and (\ref{E37}) can be interpreted as the thermodynamic speed limit of information thermodynamics based on optimal transport theory. The above findings may also explain how large the dissipation of each subsystem in the signal transduction of {\it E. coli} chemotaxis as a consequence of the evolutionary processes. 

In general, such a game-theoretic conflict of the minimum partial entropy productions arises in a more complex system such as the living organism, whereas this study only considered a simple bipartite system. For example, thermodynamics of the membrane transport~\cite{muneyuki2010allosteric,tome2015stochastic} and its information transmission~\cite{yoshida2022thermodynamic} can be discussed by using a more complex system. To understand a game-theoretic conflict in other living systems such as the membrane transport, we may need to generalize our result. Especially, how the trade-off of the minimum partial entropy productions between several subsystems emerges for complex systems is still questioned. To solve this question, we may need to consider multi-body interactions and conflicts between several subsystems. We believe that such a game-theoretic minimization of the partial entropy productions is important to explain the thermodynamic structure of the complex living systems as a consequence of evolutionary optimization.

\begin{acknowledgments}
We thank Shin-ichi Sasa, Andreas Dechant, Kohei Yoshimura, and Artemy Kolchinsky for the valuable discussions of optimal transport theory, and Kiyoshi Kanazawa for the valuable discussions of mean-field game theory.
Y.F. is supported by JSPS KAKENHI Grant No. 21J01393. S.I. is supported by JSPS KAKENHI Grant No. 19H05796, 21H01560, JST Presto Grant No. JPMJPR18M2 and UTEC-UTokyo FSI Research Grant Program.
\end{acknowledgments}

\appendix
\renewcommand{\thesection}{\Alph{section}}
\setcounter{section}{0}
\setcounter{equation}{0}
\renewcommand{\figurename}{FIG. E}
\setcounter{figure}{0}

\begin{widetext}

\section{Calculation of Nash equilibrium solution}
In this section, we concretely calculate the Nash equilibrium solution Eqs.~\eqref{E26} and \eqref{E27} under the near-equilibrium condition in linear irreversible thermodynamics~\cite{schnakenberg1976network}. By introducing the near-equilibrium condition, the minimization problem is considered within the region of linear response. Outside of the region of linear response, the minimization problem for the transition rates $\hat{W}_{X}$ and $\hat{W}_{Y}$ has a trivial solution that gives the zero entropy production without any constraints. Therefore, we only consider this problem within the non-trivial region of linear response where the non-zero entropy production can be obtained under the constraint that the Onsager coefficient is fixed.

In linear irreversible thermodynamics, (the reciprocal of) the Onsager coefficient defined as $\alpha_{z'\rightarrow z}:=(W_{z'\rightarrow z}^{\rm eq}p_{z'}^{\rm eq})^{-1}=(W_{z\rightarrow z'}^{\rm eq}p_{z}^{\rm eq})^{-1}$ leads to the linear relation $F_{e_i}^{(t)}=\alpha_{e_i}J_{e_i}^{(t)}+O(\delta W^2)$~\cite{schnakenberg1976network, yamamoto2016linear}. By using this Onsager coefficient, the partial entropy production is approximated as
\begin{align}
    \Sigma_X&=\sum_{e_i\in\mathcal{E}_X}\alpha_{e_i}  \int_{t_0}^{t_0+ \tau} dt (J_{e_i}^{(t)})^2,
    \label{EA17}\\
    \Sigma_Y&=\sum_{e_i\in\mathcal{E}_Y} \alpha_{e_i} \int_{t_0}^{t_0+ \tau} dt (J_{e_i}^{(t)})^2.
    \label{EA18}
\end{align}
by ignoring the term $O(\delta W^3)$. 

To introduce a time-averaged probability flow $\bar{J}_{z'\rightarrow z}:=(\int_{t_0}^{t_0+ \tau}dt J_{z'\rightarrow z}^{(t)})/\tau$, we solve the minimization problem of Eq.~(10a). Because $\int_{t_0}^{t_0+\tau} dt (J_{e_i}^{(t)})^2=\int_{t_0}^{t_0+\tau}dt [(\bar{J}_{e_i})^2+ ( J_{e_i}^{(t)}-\bar{J}_{e_i})^2 ]{\geq}\int_{t_0}^{t_0+\tau} dt (\bar{J}_{e_i})^2$, the lower bound on $\Sigma_X$ is given by 
\begin{align}
    \bar{\Sigma}_X (\bar{J}_X) := \tau \sum_{e_i\in\mathcal{E}_X}\alpha_{e_i} ( \bar{J}_{e_i} )^2 \leq  \Sigma_X,
    \label{EA19}
\end{align}
where $\bar{J}_X:=\{\bar{J}_{e_i}|e_i\in\mathcal{E}_X\}$ and $\bar{J}_Y:=\{\bar{J}_{e_i}|e_i\in\mathcal{E}_X\}$ are the sets of $\bar{J}_{e_i}$. The penalty $D$ is also given by the function of $\bar{J}_{e_i}$;
\begin{align}
    D&=\frac{1}{2}\sum_{i}\frac{|\delta p_{v_i}  -  \tau  \sum_j \mathsf {B}_{ij} \bar{J}_{e_j}|^2 }{p_{v_i}^{\rm f}} := \bar{D}(\bar{J}_X, \bar{J}_Y),
    \label{EA20}
\end{align}
where $\delta p_{v_i}:=p_{v_i}^{\rm i}-p_{v_i}^{\rm f}$. Thus, the minimization problem of Eq.~\eqref{E10} is rewritten as
\begin{align}
    \min_{\hat{W}_{X}}\left[\Sigma_X+\lambda_X D\right] =& \min_{\bar{J}_X}  C_X(\bar{J}_X,\bar{J}_Y),
    \label{EA21}\\
    C_X(\bar{J}_X,\bar{J}_Y) :=&\ \bar{\Sigma}_X (\bar{J}_X)  + \lambda_X \bar{D}(\bar{J}_X, \bar{J}_Y),
    \label{EA22}
\end{align}
under the near-equilibrium condition. For the subsystem $Y$, we also rewrite the minimization problem Eq.~\eqref{E11} as
\begin{align}
    \min_{\hat{W}_{Y}}\left[\Sigma_Y+\lambda_Y D\right]=&\min_{\bar{J}_{Y}}C_Y(\bar{J}_X,\bar{J}_Y)
    \label{EA23}\\
    C_Y(\bar{J}_X,\bar{J}_Y):=&\ \bar{\Sigma}_Y (\bar{J}_X)  + \lambda_Y \bar{D}(\bar{J}_X, \bar{J}_Y),
    \label{EA24}\\
    \bar{\Sigma}_Y (\bar{J}_Y):=&\  \tau  \sum_{e_i\in\mathcal{E}_Y}\alpha_{e_i} ( \bar{J}_{e_i} )^2.
    \label{EA25}
\end{align}
The above minimization problem for the player ${\cal X}$ (${\cal Y}$) is bounded by the minimization on the time-averaged probability flows $\bar{J}_X$ ($\bar{J}_Y$). 

The Nash equilibrium solution for the time-averaged probability flow $(\bar{J}_X^{\rm N},\bar{J}_Y^{\rm N})$ is given by
\begin{subequations}
\begin{empheq}[left={\empheqlbrace}]{alignat=1}
	\bar{J}_X^{\rm N}=\bar{J}_X^{*}(\bar{J}_Y^{\rm N}),
    \label{EA26}\\
	\bar{J}_Y^{\rm N}=\bar{J}_Y^{*}(\bar{J}_X^{\rm N}),
    \label{EA27}
\end{empheq}
\end{subequations}
as the fixed point of the following two solutions in the minimization problems Eqs.~(\ref{EA21}) and (\ref{EA23}),
\begin{subequations}
\begin{empheq}[left={\empheqlbrace}]{alignat=1}
	\bar{J}_X^*(\bar{J}_Y)=&{\rm arg} \min_{\bar{J}_{X}} C_X(\bar{J}_X,\bar{J}_Y),
    \label{EA28}\\
	\bar{J}_Y^*(\bar{J}_X)=&{\rm arg} \min_{\bar{J}_{Y}} C_Y(\bar{J}_X,\bar{J}_Y).
    \label{EA29}
\end{empheq}
\end{subequations}

Eqs.~(\ref{EA28})-(\ref{EA29}) show that $C_X(\bar{J}_X,\bar{J}_Y)$ and $C_Y(\bar{J}_X,\bar{J}_Y)$ take extreme values under the Nash equilibrium condition in the directions of $\bar{J}_{X}$ and $\bar{J}_{Y}$, respectively. Thus, by using
\begin{align}
\begin{split}
    C_{e_k}(\bar{J}_X,\bar{J}_Y):=&\left\{\begin{array}{ll}
        C_X(\bar{J}_X,\bar{J}_Y) & (e_k\in{\cal E}_X)\\
        C_Y(\bar{J}_X,\bar{J}_Y) & (e_k\in{\cal E}_Y)\\
    \end{array}\right.,\\
    \lambda_{e_k}:=&\left\{\begin{array}{ll}
        \lambda_X & (e_k\in{\cal E}_X)\\
        \lambda_Y & (e_k\in{\cal E}_Y)\\
    \end{array}\right.,
\end{split}
\end{align}
we calculate the extreme value conditions as
\begin{align}
\begin{split}
    &\left.\frac{\partial C_{e_k}(\bar{J}_X,\bar{J}_Y)}{\partial \bar{J}_{e_k}}\right|_{\bar{J}_X=\bar{J}_X^{\rm N}, \bar{J}_Y=\bar{J}_Y^{\rm N}} =0,\\
    &\Leftrightarrow 2\alpha_{e_k}\bar{J}_{e_k}^{\rm N}+\lambda_{e_k}\sum_i\frac{{\sf B}_{ik}}{p_{v_i}^{\rm f}}\left( \tau \sum_j{\sf B}_{ij}\bar{J}_{e_j}^{\rm N}-\delta p_{v_i}\right)=0.
\end{split}
\end{align}

By using the vector notation $(\bar{\boldsymbol{J}}^{\rm N})_i:=\bar{J}_{e_i}^{\rm N}$, these extreme value conditions are rewritten as
\begin{align}
    {\sf M}\bar{\boldsymbol{J}}^{\rm N}=&\delta\boldsymbol{q},
    \label{EA30}\\
    {\sf M}_{kj}:=&2 \frac{\alpha_{e_k}}{\lambda_{e_k}} \delta_{kj}- \tau \left(\frac{ \delta_{k \sigma(j)}- \delta_{kj}}{p_{v_k}^{\rm f}} \right) + \tau  \left(\frac{\delta_{\sigma(k) \sigma(j)}- \delta_{\sigma(k)j} }{p_{v_{\sigma(k)}}^{\rm f}}\right),
    \label{EA31}\\
    (\delta\boldsymbol{q})_k:=&\frac{\delta p_{v_{\sigma(k)}}}{p_{v_{\sigma(k)}}^{\rm f}} -\frac{\delta p_{v_k}}{p_{v_k}^{\rm f}},
    \label{EA33}
\end{align}
where we used $\mathsf{B}_{ij} = \delta_{i\sigma(j)} - \delta_{ij}$ and $\sum_{i}\mathsf{B}_{ik} \mathsf{B}_{ij}/p_{v_i}^{\rm f}=(\delta_{\sigma(k) \sigma(j)}- \delta_{j\sigma(k)})/p_{v_{\sigma(k)}}^{\rm f} - (\delta_{k\sigma(j)}- \delta_{ij} )/p_{v_{k}}^{\rm f}$.
Therefore, the Nash equilibrium solution is analytically solved by using Cramer's rule $(\bar{\boldsymbol{J}}^{\rm N})_i={\rm det}({\sf M}_i)/{\rm det}({\sf M})$. Here, ${\sf M}_i$ denotes a matrix ${\sf M}$ whose $i$-th column is replaced by $\delta\boldsymbol{q}$.

To calculate this Cramer's rule, we use the following notations for simplicity, 
\begin{align}
    &\tilde{p}_i:=p_{v_i}^{\rm f},\ \tilde{\delta}_i:=\frac{\delta p_{v_i}}{ \tau },\ \tilde{\alpha}_i:=\frac{2 \lambda_{e_i}^{-1}\alpha_{e_i}}{ \tau },
    \label{ES01}
\end{align}
and the matrix ${\sf M}$ and  ${\sf M}_i$ are given by
\begin{align}
   {\sf M}  = \tau \left(\begin{array}{cccc}
        \tilde{\alpha}_1 + \tilde{p}^{-1}_1+ \tilde{p}^{-1}_2 & -\tilde{p}^{-1}_2 &0 & -\tilde{p}^{-1}_1 \\
        -\tilde{p}^{-1}_2 & \tilde{\alpha}_2 + \tilde{p}^{-1}_2+ \tilde{p}^{-1}_3  &  -\tilde{p}^{-1}_3  & 0 \\
        0& -\tilde{p}^{-1}_3  & \tilde{\alpha}_3 + \tilde{p}^{-1}_3+ \tilde{p}^{-1}_4  &-\tilde{p}^{-1}_4 \\
        -\tilde{p}^{-1}_1 & 0 & -\tilde{p}^{-1}_4 & \tilde{\alpha}_4 + \tilde{p}^{-1}_4+ \tilde{p}^{-1}_1  \\
    \end{array}\right),
\end{align}
\begin{align}
   {\sf M}_1  = \tau \left(\begin{array}{cccc}
        \tilde{\delta}_2 \tilde{p}^{-1}_2  - \tilde{\delta}_1 \tilde{p}^{-1}_1 & -\tilde{p}^{-1}_2 &0 & -\tilde{p}^{-1}_1 \\
        \tilde{\delta}_3 \tilde{p}^{-1}_3  - \tilde{\delta}_2 \tilde{p}^{-1}_2 & \tilde{\alpha}_2 + \tilde{p}^{-1}_2+ \tilde{p}^{-1}_3  &  -\tilde{p}^{-1}_3  & 0 \\
        \tilde{\delta}_4 \tilde{p}^{-1}_4  - \tilde{\delta}_3 \tilde{p}^{-1}_3& -\tilde{p}^{-1}_3  & \tilde{\alpha}_3 + \tilde{p}^{-1}_3+ \tilde{p}^{-1}_4  &-\tilde{p}^{-1}_4 \\
        \tilde{\delta}_1 \tilde{p}^{-1}_1  - \tilde{\delta}_4 \tilde{p}^{-1}_4 & 0 & -\tilde{p}^{-1}_4 & \tilde{\alpha}_4 + \tilde{p}^{-1}_4+ \tilde{p}^{-1}_1  \\
    \end{array}\right),
\end{align}
\begin{align}
   {\sf M}_2  = \tau \left(\begin{array}{cccc}
        \tilde{\alpha}_1 + \tilde{p}^{-1}_1+ \tilde{p}^{-1}_2 & \tilde{\delta}_2 \tilde{p}^{-1}_2  - \tilde{\delta}_1 \tilde{p}^{-1}_1  &0 & -\tilde{p}^{-1}_1 \\
        -\tilde{p}^{-1}_2 & \tilde{\delta}_3 \tilde{p}^{-1}_3  - \tilde{\delta}_2 \tilde{p}^{-1}_2  &  -\tilde{p}^{-1}_3  & 0 \\
        0& \tilde{\delta}_4 \tilde{p}^{-1}_4  - \tilde{\delta}_3 \tilde{p}^{-1}_3 & \tilde{\alpha}_3 + \tilde{p}^{-1}_3+ \tilde{p}^{-1}_4  &-\tilde{p}^{-1}_4 \\
        -\tilde{p}^{-1}_1 & \tilde{\delta}_1 \tilde{p}^{-1}_1  - \tilde{\delta}_4 \tilde{p}^{-1}_4 & -\tilde{p}^{-1}_4 & \tilde{\alpha}_4 + \tilde{p}^{-1}_4+ \tilde{p}^{-1}_1  \\
    \end{array}\right),
\end{align}
\begin{align}
   {\sf M}_3  = \tau \left(\begin{array}{cccc}
        \tilde{\alpha}_1 + \tilde{p}^{-1}_1+ \tilde{p}^{-1}_2 & -\tilde{p}^{-1}_2 &\tilde{\delta}_2 \tilde{p}^{-1}_2  - \tilde{\delta}_1 \tilde{p}^{-1}_1  & -\tilde{p}^{-1}_1 \\
        -\tilde{p}^{-1}_2 & \tilde{\alpha}_2 + \tilde{p}^{-1}_2+ \tilde{p}^{-1}_3  &  \tilde{\delta}_3 \tilde{p}^{-1}_3  - \tilde{\delta}_2 \tilde{p}^{-1}_2  & 0 \\
        0& -\tilde{p}^{-1}_3  &\tilde{\delta}_4 \tilde{p}^{-1}_4  - \tilde{\delta}_3 \tilde{p}^{-1}_3  &-\tilde{p}^{-1}_4 \\
        -\tilde{p}^{-1}_1 & 0 & \tilde{\delta}_1 \tilde{p}^{-1}_1  - \tilde{\delta}_4 \tilde{p}^{-1}_4& \tilde{\alpha}_4 + \tilde{p}^{-1}_4+ \tilde{p}^{-1}_1  \\
    \end{array}\right),
\end{align}
\begin{align}
   {\sf M}_4  = \tau \left(\begin{array}{cccc}
        \tilde{\alpha}_1 + \tilde{p}^{-1}_1+ \tilde{p}^{-1}_2 & -\tilde{p}^{-1}_2 &0 & \tilde{\delta}_2 \tilde{p}^{-1}_2  - \tilde{\delta}_1 \tilde{p}^{-1}_1 \\
        -\tilde{p}^{-1}_2 & \tilde{\alpha}_2 + \tilde{p}^{-1}_2+ \tilde{p}^{-1}_3  &  -\tilde{p}^{-1}_3  & \tilde{\delta}_3 \tilde{p}^{-1}_3  - \tilde{\delta}_2 \tilde{p}^{-1}_2 \\
        0& -\tilde{p}^{-1}_3  & \tilde{\alpha}_3 + \tilde{p}^{-1}_3+ \tilde{p}^{-1}_4  &\tilde{\delta}_4 \tilde{p}^{-1}_4  - \tilde{\delta}_3 \tilde{p}^{-1}_3 \\
        -\tilde{p}^{-1}_1 & 0 & -\tilde{p}^{-1}_4 & \tilde{\delta}_1 \tilde{p}^{-1}_1  - \tilde{\delta}_4 \tilde{p}^{-1}_4  \\
    \end{array}\right).
\end{align}
The analytical solutions of $\det \mathsf{M}$ and $\det \mathsf{M_i}$ are given by
\begin{align}
   \det \sf M  =& \tau^4 \left(\sum_i \tilde{\alpha}_i \right) \left(\sum_j \tilde{p}_j \right) \left( \prod_k \tilde{p}^{-1}_k \right)+ \tau^4\left[ \sum_{i}\tilde{p}^{-1}_{i} \tilde{p}^{-1}_{\sigma(i)}\tilde{\alpha}_{\sigma^2(i)} (\tilde{\alpha}_{i}+\tilde{\alpha}_{\sigma(i)}+\tilde{\alpha}_{\sigma^3(i)}) \right] \nonumber\\
   & + \frac{1}{2}\tau^4\left[ \sum_{i}\tilde{p}_{i}^{-1} \tilde{p}_{\sigma^2(i)}^{-1}(\tilde{\alpha}_{i} +\tilde{\alpha}_{\sigma^3(i)}) (\tilde{\alpha}_{\sigma(i)}+\tilde{\alpha}_{\sigma^2(i)}) \right]  + \tau^4\left[ \sum_{i}\tilde{p}^{-1}_{i} \tilde{\alpha}_{\sigma(i)} \tilde{\alpha}_{\sigma^2(i)}  (\tilde{\alpha}_{\sigma^3(i)}+ \tilde{\alpha}_{i} )\right]
   + \tau^4 \left(\prod_k \tilde{\alpha}_k \right),
\end{align}
and 
\begin{align}
   \det \sf M_i  =& \tau^4 \left[- \tilde{\delta}_i \left \{  (\tilde{\alpha}_{\sigma(i)} +\tilde{\alpha}_{\sigma^2(i)} +\tilde{\alpha}_{\sigma^3(i)} )\tilde{p}_{\sigma(i)} + (\tilde{\alpha}_{\sigma^2(i)} +\tilde{\alpha}_{\sigma^3(i)}) \tilde{p}_{\sigma^2(i)}  + \tilde{\alpha}_{\sigma^3(i)} \tilde{p}_{\sigma^3(i)} \right\}  \right]   \left( \prod_k \tilde{p}^{-1}_k \right) \nonumber \\
   &+\tau^4 \left[\tilde{\delta}_{\sigma(i)} \left \{  (\tilde{\alpha}_{\sigma(i)} +\tilde{\alpha}_{\sigma^2(i)} +\tilde{\alpha}_{\sigma^3(i)} )\tilde{p}_{i}  + \tilde{\alpha}_{\sigma(i)} \tilde{p}_{\sigma^2(i)} + (\tilde{\alpha}_{\sigma(i)} + \tilde{\alpha}_{\sigma^2(i)} )\tilde{p}_{\sigma^3(i)} \right\}  \right]   \left( \prod_k \tilde{p}^{-1}_k \right)
    \nonumber \\
   &+\tau^4 \left[\tilde{\delta}_{\sigma^2(i)} \left \{  (\tilde{\alpha}_{\sigma^2(i)} +\tilde{\alpha}_{\sigma^3(i)}  )\tilde{p}_i -  \tilde{\alpha}_{\sigma(i)} \tilde{p}_{\sigma(i)}+  \tilde{\alpha}_{\sigma^2(i)} \tilde{p}_{\sigma^3(i)} \right\}  \right]   \left( \prod_k \tilde{p}^{-1}_k \right) \nonumber \\
   &+\tau^4 \left[\tilde{\delta}_{\sigma^3(i)} \left \{  \tilde{\alpha}_{\sigma^3(i)} \tilde{p}_i -  (\tilde{\alpha}_{\sigma(i)}  +  \tilde{\alpha}_{\sigma^2(i)}) \tilde{p}_{\sigma(i)}   -  \tilde{\alpha}_{\sigma^2(i)} \tilde{p}_{\sigma^2(i)} \right\}  \right]   \left( \prod_k \tilde{p}^{-1}_k \right) \nonumber \\
   &+\tau^4 \tilde{p}_i^{-1}\tilde{p}_{\sigma(i)}^{-1} \tilde{\alpha}_{\sigma^2(i)} (-\tilde{\alpha}_{\sigma^3(i)}  \tilde{\delta}_i +\tilde{\alpha}_{\sigma(i)}  \tilde{\delta}_{\sigma(i)}) + \tau^4\tilde{p}_i^{-1}\tilde{p}_{\sigma^2(i)}^{-1} \tilde{\alpha}_{\sigma^3(i)} (-\tilde{\alpha}_{\sigma(i)}  \tilde{\delta}_i -\tilde{\alpha}_{\sigma^2(i)}  \tilde{\delta}_i) \nonumber \\ &+\tau^4\tilde{p}_{\sigma(i)}^{-1}\tilde{p}_{\sigma^2(i)}^{-1} \tilde{\alpha}_{\sigma^3(i)} (\tilde{\alpha}_{\sigma(i)}  \tilde{\delta}_{\sigma(i)} +\tilde{\alpha}_{\sigma^2(i)}  \tilde{\delta}_{\sigma(i)} + \tilde{\alpha}_{\sigma^2(i)}  \tilde{\delta}_{\sigma^2(i)} )  -\tau^4\tilde{p}_i^{-1}\tilde{p}_{\sigma^3(i)}^{-1} \tilde{\alpha}_{\sigma(i)} (\tilde{\alpha}_{\sigma^2(i)}  \tilde{\delta}_i +\tilde{\alpha}_{\sigma^3(i)}  \tilde{\delta}_i + \tilde{\alpha}_{\sigma^2(i)}  \tilde{\delta}_{\sigma^3(i)} ) \nonumber \\
   &+ \tau^4\tilde{p}_{\sigma(i)}^{-1}\tilde{p}_{\sigma^3(i)}^{-1} \tilde{\alpha}_{\sigma(i)} (\tilde{\alpha}_{\sigma^2(i)}   +\tilde{\alpha}_{\sigma^3(i)} ) \tilde{\delta}_{\sigma(i)}
   +\tau^4 \tilde{\alpha}_{\sigma(i)} \tilde{\alpha}_{\sigma^2(i)} \tilde{\alpha}_{\sigma^3(i)} ( -   \tilde{p}_i^{-1} \tilde{\delta}_i+\tilde{p}_{\sigma(i)}^{-1} \tilde{\delta}_{\sigma(i)}).
\end{align}
Thus, $(\bar{\boldsymbol{J}}^{\rm N})_i={\rm det}({\sf M}_i)/{\rm det}({\sf M})$ is analytically obtained. 

In particular, we consider $(\bar{\boldsymbol{J}}^{\rm N})_i$ in the limit $\lambda_X\rightarrow\infty$ for the fixed $r = \lambda_{X}/\lambda_{Y}$. Because $\tilde{\alpha}_i = O(\lambda_{e_i}^{-1})=O(\lambda_X^{-1})$, the second-order, the third-order and the fourth-order of $\tilde{\alpha}_i$ give $O(\lambda_X^{-2})$. Thus, we obtain
\begin{align}
   \frac{\det \mathsf{M} \left( \prod_k \tilde{p}_k \right) }{\tau^4}  =  \sum_i \tilde{\alpha}_i \ + O(\lambda_X^{-2}),
\end{align}
and
\begin{eqnarray}
    \frac{{\rm det}{\sf M}_i \left( \prod_k \tilde{p}_k \right)}{\tau^4}&=&  - \tilde{\delta}_i \left \{  (\tilde{\alpha}_{\sigma(i)} +\tilde{\alpha}_{\sigma^2(i)} +\tilde{\alpha}_{\sigma^3(i)} )\tilde{p}_{\sigma(i)} + (\tilde{\alpha}_{\sigma^2(i)} +\tilde{\alpha}_{\sigma^3(i)}) \tilde{p}_{\sigma^2(i)}  + \tilde{\alpha}_{\sigma^3(i)} \tilde{p}_{\sigma^3(i)} \right\}   \nonumber \\
   &&+\tilde{\delta}_{\sigma(i)} \left \{  (\tilde{\alpha}_{\sigma(i)} +\tilde{\alpha}_{\sigma^2(i)} +\tilde{\alpha}_{\sigma^3(i)} )\tilde{p}_{i}  + \tilde{\alpha}_{\sigma(i)} \tilde{p}_{\sigma^2(i)} + (\tilde{\alpha}_{\sigma(i)} + \tilde{\alpha}_{\sigma^2(i)} )\tilde{p}_{\sigma^3(i)} \right\}
    \nonumber \\
   &&+\tilde{\delta}_{\sigma^2(i)} \left \{  (\tilde{\alpha}_{\sigma^2(i)} +\tilde{\alpha}_{\sigma^3(i)}  )\tilde{p}_i -  \tilde{\alpha}_{\sigma(i)} \tilde{p}_{\sigma(i)}+  \tilde{\alpha}_{\sigma^2(i)} \tilde{p}_{\sigma^3(i)} \right\}   \nonumber \\
   &&+\tilde{\delta}_{\sigma^3(i)} \left \{  \tilde{\alpha}_{\sigma^3(i)} \tilde{p}_i -  (\tilde{\alpha}_{\sigma(i)}  +  \tilde{\alpha}_{\sigma^2(i)}) \tilde{p}_{\sigma(i)}   -  \tilde{\alpha}_{\sigma^2(i)} \tilde{p}_{\sigma^2(i)} \right\}  + O(\lambda_X^{-2})
    \nonumber\\
    &=&  \tb_{\sigma(i)} (-\tp_{\sigma(i)} \td_i + \tp_i  \td_{\sigma(i)} + \tp_{\sigma^2(i)}  \td_{\sigma(i)}  +\tp_{\sigma^3(i)} \td_{\sigma(i)}   - \tp_{\sigma(i)}\td_{\sigma^2(i)} -  \tp_{\sigma(i)} \td_{\sigma^3(i)})   \nonumber \\
    &&+\tb_{\sigma^2(i)} (-\tp_{\sigma(i)} \td_i -\tp_{\sigma^2(i)}  \td_{i} + \tp_i  \td_{\sigma(i)} +\tp_{\sigma^3(i)} \td_{\sigma(i)} +\tp_{i} \td_{\sigma^2(i)} +\tp_{\sigma^3(i)} \td_{\sigma^2(i)} - \tp_{\sigma(i)} \td_{\sigma^3(i)} - \tp_{\sigma^2(i)} \td_{\sigma^3(i)}  )   \nonumber \\
    &&+\tb_{\sigma^3(i)} (-\tp_{\sigma(i)} \td_i - \tp_{\sigma^2(i)} \td_i -\tp_{\sigma^3(i)} \td_i + \tp_{i} \td_{\sigma(i)}  + \tp_{i} \td_{\sigma^2(i)} +   \tp_{i} \td_{\sigma^3(i)})   + O(\lambda_X^{-2})
    \nonumber\\
    &=& \left \{\tb_{\sigma(i)} \td_{\sigma(i)}+\tb_{\sigma^2(i)} (\td_{\sigma(i)}+\td_{\sigma^2(i)}) -\tb_{\sigma^3(i)}\td_i \right\} \left( \sum_j \tp_j \right) \nonumber \\
    && + \left\{ - \tb_{\sigma(i)}\tp_{\sigma(i)} - \tb_{\sigma^2(i)}(\tp_{\sigma(i)} +\tp_{\sigma^2(i)}) + \tb_{\sigma^3(i)}\tp_i \right\} \left(\sum_j \td_j \right)   + O(\lambda_X^{-2})
    \nonumber\\
    &=&-\tb_{\sigma^3(i)}\td_i + ( \tb_{\sigma(i)} +\tb_{\sigma^2(i)} ) \td_{\sigma(i)}+\tb_{\sigma^2(i)} \td_{\sigma^2(i)} + O(\lambda_X^{-2}).
    \label{ES05}
\end{eqnarray}
where we used $\tp_i + \tp_{\sigma(i)} +\tp_{\sigma^2(i)} +\tp_{\sigma^3(i)}=\sum_{j} \tp_j =1 $ and $\td_i + \td_{\sigma(i)} +\td_{\sigma^2(i)} +\td_{\sigma^3(i)} = \sum_j \td_j = \sum_j (p^{\rm i}_{v_j} - p^{\rm f}_{v_j}) /\tau=0$. By using these equations and substituting the definitions of Eqs.~(\ref{ES01}) into Eq.~(\ref{ES05}), we compute the flow $\bar{J}_{e_i}^{\rm N}$ as
\begin{align}
    \bar{J}_{e_i}^{\rm N}&=\lim_{\lambda_X \to \infty | \lambda_X/\lambda_Y = r}  \frac{{\rm det}{\sf M}_i}{{\rm det}{\sf M}}
    \nonumber\\
    &=\lim_{\lambda_X \to \infty | \lambda_X/\lambda_Y = r}  \left( \frac{{\rm det}{\sf M}_i \left( \prod_k \tilde{p}_k \right)}{\tau^4} \right) \left( \frac{{\rm det}{\sf M}\left( \prod_k \tilde{p}_k \right)}{\tau^4}\right)^{-1}
    \nonumber\\&=\lim_{\lambda_X \to \infty | \lambda_X/\lambda_Y = r}   \frac{-\tb_{\sigma^3(i)}\td_i + ( \tb_{\sigma(i)} +\tb_{\sigma^2(i)} ) \td_{\sigma(i)}+\tb_{\sigma^2(i)} \td_{\sigma^2(i)}  + O(\lambda_X^{-2})}{\sum_j\tb_j  + O(\lambda_X^{-2})} \nonumber\\
    &=\lim_{\lambda_X \to \infty | \lambda_X/\lambda_Y = r}   \frac{- (\lambda_{e_{\sigma^3(i)}}^{-1}\alpha_{e_{\sigma^3(i)}} ) \delta p_{v_i} + ( \lambda_{e_{\sigma(i)}}^{-1}\alpha_{e_{\sigma(i)}}  +\lambda_{e_{\sigma^2(i)}}^{-1}\alpha_{e_{\sigma^2(i)}} )  \delta p_{v_{\sigma(i)}} +(\lambda_{e_{\sigma^2(i)}}^{-1}\alpha_{e_{\sigma^2(i)}} )\delta p_{v_{\sigma^2(i)}}   + O(\lambda_X^{-2})}{ \tau(\sum_j \lambda_{e_j}^{-1}\alpha_{e_j} )  + O(\lambda_X^{-2})}.
    \label{ES06}
\end{align}

By using $\alpha_X = \alpha_{e_1} +\alpha_{e_3}$, $\alpha_Y = \alpha_{e_2} +\alpha_{e_4}$, $\lambda_{e_1}=\lambda_{e_3} = \lambda_X $ and $\lambda_{e_2}=\lambda_{e_4} = \lambda_Y =  \lambda_X/r$, we concretely compute the flows as
\begin{align}
    \bar{J}_{e_1}^{\rm N}&=\lim_{\lambda_X \to \infty | \lambda_X/\lambda_Y = r}   \frac{-r\lambda_X^{-1}\alpha_{e_4}\delta p_{v_1}+(r \lambda_X^{-1}\alpha_{e_2}+\lambda_X^{-1}\alpha_{e_3})\delta p_{v_2}+\lambda_X^{-1}\alpha_{e_3}\delta p_{v_3}  + O(\lambda_X^{-2})}{\tau (\lambda_X^{-1}\alpha_{e_1}+r \lambda_X^{-1}\alpha_{e_2}+\lambda_X^{-1}\alpha_{e_3}+r \lambda_X^{-1}\alpha_{e_4}) + O(\lambda_X^{-2}) }
    \nonumber\\
    &=\frac{-r\alpha_{e_4}\delta p_{v_1}+(r\alpha_{e_2}+\alpha_{e_3})\delta p_{v_2}+\alpha_{e_3}\delta p_{v_3}}{\tau (\alpha_X+r\alpha_Y)},
    \label{ES07}\\
    \bar{J}_{e_2}^{\rm N}&=\lim_{\lambda_X \to \infty | \lambda_X/\lambda_Y = r}  \frac{-\lambda_X^{-1}\alpha_{e_1}\delta p_{v_2}+(\lambda_X^{-1}\alpha_{e_3}+ r \lambda_X^{-1}\alpha_{e_4})\delta p_{v_3}+r \lambda_X^{-1}\alpha_{e_4}\delta p_{v_4} + O(\lambda_X^{-2})}{ \tau(\lambda_X^{-1}\alpha_{e_1}+ r \lambda_X^{-1}\alpha_{e_2}+\lambda_X^{-1}\alpha_{e_3}+ r \lambda_X^{-1}\alpha_{e_4}) + O(\lambda_X^{-2})}
    \nonumber\\
    &=\frac{-\alpha_{e_1}\delta p_{v_2}+(\alpha_{e_3}+r\alpha_{e_4})\delta p_{v_3}+r\alpha_{e_4}\delta p_{v_4}}{\tau (\alpha_X+r\alpha_Y)},
    \label{ES08}\\
    \bar{J}_{e_3}^{\rm N}&= \lim_{\lambda_X \to \infty | \lambda_X/\lambda_Y = r}  \frac{- r \lambda_X^{-1}\alpha_{e_2}\delta p_{v_3}+(r \lambda_X^{-1}\alpha_{e_4}+\lambda_X^{-1}\alpha_{e_1})\delta p_{v_4}+\lambda_X^{-1}\alpha_{e_1}\delta p_{v_1} + O(\lambda_X^{-2})}{\tau (\lambda_X^{-1}\alpha_{e_1}+ r \lambda_X^{-1}\alpha_{e_2}+\lambda_X^{-1}\alpha_{e_3}+ r \lambda_X^{-1}\alpha_{e_4}) + O(\lambda_X^{-2})}
    \nonumber\\
    &=\frac{-r\alpha_{e_2}\delta p_{v_3}+(r\alpha_{e_4}+\alpha_{e_1})\delta p_{v_4}+\alpha_{e_1}\delta p_{v_1}}{\tau (\alpha_X+r\alpha_Y)},
    \label{ES09}\\
    \bar{J}_{e_4}^{\rm N}&= \lim_{\lambda_X \to \infty | \lambda_X/\lambda_Y = r}  \frac{-\lambda_X^{-1}\alpha_{e_3}\delta p_{v_4}+(\lambda_X^{-1}\alpha_{e_1}+r \lambda_X^{-1}\alpha_{e_2})\delta p_{v_1}+ r \lambda_X^{-1}\alpha_{e_2}\delta p_{v_2}+ O(\lambda_X^{-2})}{\tau (\lambda_X^{-1}\alpha_{e_1}+r \lambda_X^{-1}\alpha_{e_2}+\lambda_X^{-1}\alpha_{e_3}+r \lambda_X^{-1}\alpha_{e_4})+ O(\lambda_X^{-2})}
    \nonumber\\
    &=\frac{-\alpha_{e_3}\delta p_{v_4}+(\alpha_{e_1}+r\alpha_{e_2})\delta p_{v_1}+r\alpha_{e_2}\delta p_{v_2}}{\tau (\alpha_X+r\alpha_Y)},
    \label{ES10}
\end{align}
which corresponds to Eqs.~\eqref{E34} and \eqref{E35},
\begin{align}
   \bar{J}_{e_i}^{\rm N} =& \frac{-r \alpha_{e_{\sigma^3(i)}}\delta p_{v_{i}} +(r \alpha_{e_{\sigma(i)}}+\alpha_{e_{\sigma^2(i)}})\delta p_{v_{\sigma(i)}} +\alpha_{e_{\sigma^2(i)}}\delta p_{v_{\sigma^2(i)}} }{{\tau(\alpha_X + r \alpha_Y})} ,
   \label{EA34}
\end{align}
for $e_i\in{\cal E}_X$, and
\begin{align}
   \bar{J}_{e_i}^{\rm N} =& \frac{ -\alpha_{e_{\sigma^3(i)}}\delta p_{v_{i}} +(\alpha_{e_{\sigma(i)}}+r\alpha_{e_{\sigma^2(i)}})\delta p_{v_{\sigma(i)}} +r\alpha_{e_{\sigma^2(i)}}\delta p_{v_{\sigma^2(i)}}}{{\tau (\alpha_X + r \alpha_Y})},
   \label{EA35}
\end{align}
for $e_i\in{\cal E}_Y$.

\section{Analytical expressions of $\bar{\boldsymbol{\mathcal{J}}}^{\rm N}$}
In this section, we calculate $\bar{\boldsymbol{\mathcal{J}}}^{\rm N}$ in the limit $\lambda_X \to \infty$ with the fixed ratio $r$. From the definition of $\bar{\boldsymbol{\mathcal{J}}}^{\rm N}$ in Eqs.~\eqref{E39}-\eqref{E41}, we obtain
\begin{align}
    \bar{\mathcal{J}}_{X}^{\rm N}&=\bar{J}_{e_1}^{\rm N}-\bar{J}_{e_3}^{\rm N}
    \nonumber\\
    &=\frac{-(\alpha_{e_1}+r\alpha_{e_4})(\delta p_{v_1}+\delta p_{v_4})+(r\alpha_{e_2}+\alpha_{e_3})(\delta p_{v_2}+\delta p_{v_3})}{\tau(\alpha_X+r\alpha_Y)}
    \nonumber\\
    &=\frac{(\alpha_X+r\alpha_Y)(\delta p_{v_2}+\delta p_{v_3})-(\alpha_{e_1}+r\alpha_{e_4}) \left( \sum_i\delta p_{v_i} \right)}{\tau (\alpha_X+r\alpha_Y)}
    \nonumber\\
    &=\frac{\delta p_{v_2}+\delta p_{v_3}}{ \tau },
    \label{ES11}
    \end{align}
    \begin{align}
    \bar{\mathcal{J}}_{Y}^{\rm N}&=\bar{J}_{e_2}^{\rm N}-\bar{J}_{e_4}^{\rm N}
    \nonumber\\
    &= \frac{(\alpha_{e_3}+r\alpha_{e_4})(\delta p_{v_3}+\delta p_{v_4})-(\alpha_{e_1}+r\alpha_{e_2})(\delta p_{v_1}+\delta p_{v_2})}{\tau (\alpha_X+r\alpha_Y)}
    \nonumber\\
    &=\frac{(\alpha_X+r\alpha_Y)(\delta p_{v_3}+\delta p_{v_4})-(\alpha_{e_1}+r\alpha_{e_2})\left( \sum_i\delta p_{v_i} \right)}{ \tau (\alpha_X+r\alpha_Y)}
    \nonumber\\
    &=\frac{\delta p_{v_3}+\delta p_{v_4}}{ \tau },
    \label{ES12}
        \end{align}
    \begin{align}
    \bar{\mathcal{J}}_{XY}^{\rm N}&=\frac{\alpha_{e_1}}{\alpha_X}\bar{J}_{e_1}^{\rm N}-\frac{\alpha_{e_2}}{\alpha_Y}\bar{J}_{e_2}^{\rm N}+\frac{\alpha_{e_3}}{\alpha_X}\bar{J}_{e_3}^{\rm N}-\frac{\alpha_{e_4}}{\alpha_Y}\bar{J}_{e_4}^{\rm N}
    \nonumber\\
    &=\frac{1}{  \alpha_X\alpha_Y}\{-\alpha_{e_4}\alpha_{e_1}(\bar{J}_{e_4}^{\rm N}-\bar{J}_{e_1}^{\rm N})+\alpha_{e_1}\alpha_{e_2}(\bar{J}_{e_1}^{\rm N}-\bar{J}_{e_2}^{\rm N})-\alpha_{e_2}\alpha_{e_3}(\bar{J}_{e_2}^{\rm N}-\bar{J}_{e_3}^{\rm N})+\alpha_{e_3}\alpha_{e_4}(\bar{J}_{e_3}^{\rm N}-\bar{J}_{e_4}^{\rm N})\}
    \nonumber\\
    &=\frac{-\alpha_{e_4}\alpha_{e_1}\delta p_{v_1}+\alpha_{e_1}\alpha_{e_2}\delta p_{v_2}-\alpha_{e_2}\alpha_{e_3}\delta p_{v_3}+\alpha_{e_3}\alpha_{e_4}\delta p_{v_4}}{ \tau  \alpha_X\alpha_Y},
    \label{ES13}
\end{align}
where we used $\sum_i \delta p_{v_i}=0$, $\alpha_X = \alpha_{e_1} +\alpha_{e_3}$,  $\alpha_Y = \alpha_{e_2} +\alpha_{e_4}$, and $ \bar{J}_{e_i}^{\rm N} - \bar{J}_{e_{\sigma(i)}}^{\rm N} = \delta p_{v_{\sigma(i)}}/\tau$. Here, $ \bar{J}_{e_i}^{\rm N} - \bar{J}_{e_{\sigma(i)}}^{\rm N} = \delta p_{v_{\sigma(i)}}/\tau$ can be obtained from Eq.~(\ref{ES06}) as follows,
\begin{eqnarray}
\bar{J}_{e_i}^{\rm N} - \bar{J}_{e_{\sigma(i)}}^{\rm N}&=& \frac{- (\lambda_{e_{\sigma^3(i)}}^{-1}\alpha_{e_{\sigma^3(i)}} ) \delta p_{v_i} + ( \lambda_{e_{\sigma(i)}}^{-1}\alpha_{e_{\sigma(i)}}  +\lambda_{e_{\sigma^2(i)}}^{-1}\alpha_{e_{\sigma^2(i)}} )  \delta p_{v_{\sigma(i)}} +(\lambda_{e_{\sigma^2(i)}}^{-1}\alpha_{e_{\sigma^2(i)}} )\delta p_{v_{\sigma^2(i)}}   }{ \tau(\sum_i \lambda_{e_i}^{-1}\alpha_{e_i} ) } \nonumber \\
&&- \frac{- (\lambda_{e_{i}}^{-1}\alpha_{e_{i}} ) \delta p_{v_{\sigma(i)}} + ( \lambda_{e_{\sigma^2(i)}}^{-1}\alpha_{e_{\sigma^2(i)}}  +\lambda_{e_{\sigma^3(i)}}^{-1}\alpha_{e_{\sigma^3(i)}} )  \delta p_{v_{\sigma^2(i)}} +(\lambda_{e_{\sigma^3(i)}}^{-1}\alpha_{e_{\sigma^3(i)}} )\delta p_{v_{\sigma^3(i)}}   }{ \tau(\sum_j \lambda_{e_j}^{-1}\alpha_{e_j} ) } \nonumber \\
&=& \frac{(\sum_j \lambda_{e_j}^{-1}\alpha_{e_j} )\delta p_{v_{\sigma(i)}}  }{ \tau(\sum_j \lambda_{e_j}^{-1}\alpha_{e_j} ) } \nonumber \\
&=& \frac{\delta p_{v_{\sigma(i)}}}{\tau },
\end{eqnarray}
where we used $\sum_i \delta p_{v_i} =\delta p_{v_i} +\delta p_{v_{\sigma(i)}}+\delta p_{v_{\sigma^2(i)}}+\delta p_{v_{\sigma^3(i)}} =0$. This result $ \bar{J}_{e_i}^{\rm N} - \bar{J}_{e_{\sigma(i)}}^{\rm N}= \delta p_{v_{\sigma(i)}}/\tau $ is consistent with $\delta \boldsymbol{p}= \tau \mathsf{B} \boldsymbol{\bar{J}}^{\rm N}$ where $(\delta \boldsymbol{p})_i = \delta p_{v_i}$. We can see that these flows are determined only by the state transition $\delta\boldsymbol{p}$ and independent of $r$.

We next discuss $\bar{\mathcal{J}}_{\rm rot}^{\rm N}$. To calculate $\bar{\mathcal{J}}_{\rm rot}^{\rm N}$, we obtain $\left.\bar{J}_{e_i}^{\rm N}\right|_{r=1}$ and $\bar{J}_{e_i}^{\rm N}-\left.\bar{J}_{e_i}^{\rm N}\right|_{r=1}$ from Eqs.~\eqref{ES07}-\eqref{ES10} as follows,
\begin{align}
    \left.\bar{J}_{e_i}^{\rm N}\right|_{r=1}&=\frac{-\alpha_{e_{\sigma^3(i)}}\delta p_{v_i}+(\alpha_{e_{\sigma(i)}}+\alpha_{e_{\sigma^2(i)}})\delta p_{v_{\sigma(i)}}+\alpha_{e_{\sigma^2(i)}}\delta p_{e_{\sigma^2(i)}}}{\tau(\alpha_X+\alpha_Y)},
    \label{ES14}
    \end{align}
    \begin{align}
    \bar{J}_{e_1}^{\rm N}-\left.\bar{J}_{e_1}^{\rm N}\right|_{r=1}&= \frac{-r\alpha_{e_4}\delta p_{v_1}+(r\alpha_{e_2}+\alpha_{e_3})\delta p_{v_2}+\alpha_{e_3}\delta p_{v_3}}{ \tau(\alpha_X+r\alpha_Y)}-\frac{-\alpha_{e_4}\delta p_{v_1}+(\alpha_{e_2}+\alpha_{e_3})\delta p_{v_2}+\alpha_{e_3}\delta p_{v_3}}{ \tau(\alpha_X+\alpha_Y)}
    \nonumber\\
    &=\frac{ -(r-1)\alpha_X\alpha_{e_4}\delta p_{v_1}+(r-1)\alpha_X\alpha_{e_2}\delta p_{v_2} -(r-1)\alpha_Y \alpha_{e_3}\delta p_{v_2}  -(r-1)\alpha_Y\alpha_{e_3}\delta p_{v_3} }{ \tau(\alpha_X+r\alpha_Y)(\alpha_X+\alpha_Y)}
    \nonumber\\
    &=\frac{(r-1)\left\{-\alpha_{e_4}\alpha_{e_1}\delta p_{v_1}+\alpha_{e_1}\alpha_{e_2}\delta p_{v_2}-\alpha_{e_2}\alpha_{e_3}\delta p_{v_3}+\alpha_{e_3}\alpha_{e_4}\delta p_{v_4}-\alpha_{e_3}\alpha_{e_4} \left( \sum_i \delta p_{v_i} \right) \right\}}{ \tau(\alpha_X+\alpha_Y)(\alpha_X+r\alpha_Y) }
    \nonumber\\
    &=\frac{(r-1)\alpha_X\alpha_Y}{(\alpha_X+\alpha_Y)(\alpha_X+r\alpha_Y)}\bar{\mathcal{J}}_{XY}^{\rm N},
    \label{ES15}
    \end{align}
    \begin{align}
    \bar{J}_{e_2}^{\rm N}-\left.\bar{J}_{e_2}^{\rm N}\right|_{r=1}&= \frac{-\alpha_{e_1}\delta p_{v_2}+(\alpha_{e_3}+r\alpha_{e_4})\delta p_{v_3}+r\alpha_{e_4}\delta p_{v_4}}{ \tau(\alpha_X+r\alpha_Y)}-\frac{-\alpha_{e_1}\delta p_{v_2}+(\alpha_{e_3}+\alpha_{e_4})\delta p_{v_3}+\alpha_{e_4}\delta p_{v_4}}{ \tau(\alpha_X+\alpha_Y)}
    \nonumber\\
    &=\frac{ (r-1)\alpha_X\alpha_{e_4}\delta p_{v_3}+(r-1)\alpha_X\alpha_{e_4}\delta p_{v_4} +(r-1)\alpha_Y \alpha_{e_1}\delta p_{v_2}  -(r-1)\alpha_Y\alpha_{e_3}\delta p_{v_3} }{ \tau(\alpha_X+r\alpha_Y)(\alpha_X+\alpha_Y)}
    \nonumber\\
    &=\frac{(r-1) \left\{-\alpha_{e_4}\alpha_{e_1}\delta p_{v_1}+\alpha_{e_1}\alpha_{e_2}\delta p_{v_2}-\alpha_{e_2}\alpha_{e_3}\delta p_{v_3}+\alpha_{e_3}\alpha_{e_4}\delta p_{v_4}+\alpha_{e_4}\alpha_{e_1}\left( \sum_i \delta p_{v_i} \right) \right\}}{ \tau (\alpha_X+\alpha_Y)(\alpha_X+r\alpha_Y)}
    \nonumber\\
    &=\frac{(r-1)\alpha_X\alpha_Y}{(\alpha_X+\alpha_Y)(\alpha_X+r\alpha_Y)}\bar{\mathcal{J}}_{XY}^{\rm N},
    \label{ES16}
    \end{align}
    \begin{align}
    \bar{J}_{e_3}^{\rm N}-\left.\bar{J}_{e_3}^{\rm N}\right|_{r=1}&= \frac{-r\alpha_{e_2}\delta p_{v_3}+(r\alpha_{e_4}+\alpha_{e_1})\delta p_{v_4}+\alpha_{e_1}\delta p_{v_1}}{ \tau (\alpha_X+r\alpha_Y)}-\frac{-\alpha_{e_2}\delta p_{v_3}+(\alpha_{e_4}+\alpha_{e_1})\delta p_{v_4}+\alpha_{e_1}\delta p_{v_1}}{ \tau (\alpha_X+\alpha_Y)}
    \nonumber\\
     &=\frac{ -(r-1)\alpha_X\alpha_{e_2}\delta p_{v_3}+(r-1)\alpha_X\alpha_{e_4}\delta p_{v_4} -(r-1)\alpha_Y \alpha_{e_1}\delta p_{v_4}  -(r-1)\alpha_Y\alpha_{e_1}\delta p_{v_1} }{ \tau(\alpha_X+r\alpha_Y)(\alpha_X+\alpha_Y)}
    \nonumber\\
    &=\frac{(r-1)\{-\alpha_{e_4}\alpha_{e_1}\delta p_{v_1}+\alpha_{e_1}\alpha_{e_2}\delta p_{v_2}-\alpha_{e_2}\alpha_{e_3}\delta p_{v_3}+\alpha_{e_3}\alpha_{e_4}\delta p_{v_4}-\alpha_{e_1}\alpha_{e_2} \left(\sum_i\delta p_{v_i} \right) \}}{ \tau (\alpha_X+\alpha_Y)(\alpha_X+r\alpha_Y)}
    \nonumber\\
    &=\frac{(r-1)\alpha_X\alpha_Y}{(\alpha_X+\alpha_Y)(\alpha_X+r\alpha_Y)}\bar{\mathcal{J}}_{XY}^{\rm N},
    \label{ES17}
    \end{align}
    \begin{align}
    \bar{J}_{e_4}^{\rm N}-\left.\bar{J}_{e_4}^{\rm N}\right|_{r=1}&= \frac{-\alpha_{e_3}\delta p_{v_4}+(\alpha_{e_1}+r\alpha_{e_2})\delta p_{v_1}+r\alpha_{e_2}\delta p_{v_2}}{ \tau (\alpha_X+r\alpha_Y)}-\frac{-\alpha_{e_3}\delta p_{v_4}+(\alpha_{e_1}+\alpha_{e_2})\delta p_{v_1}+\alpha_{e_2}\delta p_{v_2}}{ \tau (\alpha_X+\alpha_Y)}
    \nonumber\\
    &=\frac{ (r-1)\alpha_X\alpha_{e_2}\delta p_{v_1}+(r-1)\alpha_X\alpha_{e_2}\delta p_{v_2} +(r-1)\alpha_Y \alpha_{e_3}\delta p_{v_4}  -(r-1)\alpha_Y\alpha_{e_1}\delta p_{v_1} }{ \tau(\alpha_X+r\alpha_Y)(\alpha_X+\alpha_Y)}
    \nonumber\\
    &=\frac{(r-1)\{-\alpha_{e_4}\alpha_{e_1}\delta p_{v_1}+\alpha_{e_1}\alpha_{e_2}\delta p_{v_2}-\alpha_{e_2}\alpha_{e_3}\delta p_{v_3}+\alpha_{e_3}\alpha_{e_4}\delta p_{v_4}+\alpha_{e_2}\alpha_{e_3} \left(\sum_i\delta p_{v_i}\right)\}}{\tau(\alpha_X+\alpha_Y)(\alpha_X+r\alpha_Y)}
    \nonumber\\
    &=\frac{(r-1)\alpha_X\alpha_Y}{(\alpha_X+\alpha_Y)(\alpha_X+r\alpha_Y)}\bar{\mathcal{J}}_{XY}^{\rm N},
    \label{ES18}
\end{align}
where we used $\alpha_X = \alpha_{e_1} +\alpha_{e_3}$, $\alpha_Y = \alpha_{e_2} +\alpha_{e_4}$ and $\sum_i \delta p_{v_i}=0$. From these equations, we obtain
\begin{align}
    \bar{\mathcal{J}}_{\rm rot}^{\rm N}&=\frac{1}{4}\sum_i\bar{J}_{e_i}^{\rm N}
    \nonumber\\
    &=\frac{1}{4}\sum_i(\bar{J}_{e_i}^{\rm N}-\left.\bar{J}_{e_i}^{\rm N}\right|_{r=1})+\frac{1}{4}\sum_i\left.\bar{J}_{e_i}^{\rm N}\right|_{r=1}
    \nonumber\\
    &=\frac{(r-1)\alpha_{X}\alpha_{Y}}{(\alpha_{X}+\alpha_{Y})(\alpha_{X}+r\alpha_{Y})}\bar{\mathcal{J}}_{XY}^{\rm N}+\frac{\sum_i(2\alpha_{e_i} +\alpha_{e_{\sigma(i)}}-\alpha_{e_{\sigma^3(i)}})\delta p_{v_i}}{ 4\tau(\alpha_X+\alpha_Y)},
    \label{ES19}
\end{align}
where we used $\sum_i \alpha_{e_{\sigma^2(i)}}\delta p_{v_{\sigma(i)}} = \sum_i \alpha_{e_{\sigma(i)}}\delta p_{i}$ and $\sum_i \alpha_{e_{i}}\delta p_{v_i} = \sum_i \alpha_{e_{\sigma(i)}}\delta p_{v_{\sigma(i)}}= \sum_i \alpha_{e_{\sigma^2(i)}}\delta p_{v_{\sigma^2(i)}}$. Here, we can see that $\bar{\mathcal{J}}_{\rm rot}^{\rm N}$ depends on $r$ and a strength of this dependence is determined by $\bar{\mathcal{J}}_{XY}^{\rm N}$.  This $\bar{\mathcal{J}}_{\rm rot}^{\rm N}$ means a counterclockwise rotation of probability flows from the definition, and $\bar{\mathcal{J}}_{\rm rot}^{\rm N}$ does not contribute to the time evolution of probability distribution $\delta\boldsymbol{p}$. Because ${\rm det}{\sf T} = 1$, ${\sf T}$ has the inverse matrix
\begin{align}
   {\sf T}^{-1}  = \left(\begin{array}{cccc}
        \frac{1}{4} + \frac{\alpha_{e_3}}{2 \alpha_X} & \frac{1}{4} - \frac{\alpha_{e_4}}{2\alpha_Y} & \frac{1}{2} & 1 \\
        \frac{1}{4} - \frac{\alpha_{e_3}}{2 \alpha_X}& \frac{1}{4} + \frac{\alpha_{e_4}}{2\alpha_Y}  & -\frac{1}{2}  & 1 \\
        -\frac{1}{4} - \frac{\alpha_{e_1}}{2 \alpha_X}& -\frac{1}{4} + \frac{\alpha_{e_2}}{2\alpha_Y}  & \frac{1}{2}  & 1 \\
        -\frac{1}{4} + \frac{\alpha_{e_1}}{2 \alpha_X}& -\frac{1}{4} - \frac{\alpha_{e_2}}{2\alpha_Y} & -\frac{1}{2}  & 1 \\
    \end{array}\right),
\end{align}
and $\bar{\boldsymbol{J}}^{\rm N} =  {\sf T}^{-1} \bar{\boldsymbol{\mathcal{J}}}^{\rm N}$. By using
\begin{align}
   {\sf B}  = \left(\begin{array}{cccc}
        -1 & 0 &0 & 1 \\
        1& -1  &0  & 0 \\
        0& 1  & -1 & 0\\
        0& 0 & 1 & -1 \\
    \end{array}\right),
\end{align}
$\bar{\boldsymbol{J}}^{\rm N} =  {\sf T}^{-1} \bar{\boldsymbol{\mathcal{J}}}^{\rm N}$, and $\delta \boldsymbol{p} = \tau {\sf B}\bar{\boldsymbol{J}}^{\rm N}$, 
we can describe $\delta\boldsymbol{p}$ as the linear function of $\bar{\boldsymbol{\mathcal{J}}}^{\rm N}$
\begin{align}
   \delta\boldsymbol{p} =&\tau \mathsf{B} {\sf T}^{-1} \bar{\boldsymbol{\mathcal{J}}}^{\rm N} = \tau  \left(\begin{array}{cccc}
        -\frac{\alpha_{e_3}}{\alpha_{X}} & -\frac{\alpha_{e_2}}{\alpha_{Y}} & -1 & 0 \\
        \frac{\alpha_{e_3}}{\alpha_{X}} & -\frac{\alpha_{e_4}}{\alpha_{Y}} & 1 & 0 \\
        \frac{\alpha_{e_1}}{\alpha_{X}} & \frac{\alpha_{e_4}}{\alpha_{Y}} & -1 & 0 \\
        -\frac{\alpha_{e_1}}{\alpha_{X}} & \frac{\alpha_{e_2}}{\alpha_{Y}} & 1 & 0 \\
    \end{array}\right) \left(\begin{array}{c}
        \bar{\mathcal{J}}_{X}^{\rm N}\\
        \bar{\mathcal{J}}_{Y}^{\rm N}\\
        \bar{\mathcal{J}}_{XY}^{\rm N}\\
        \bar{\mathcal{J}}_{\rm rot}^{\rm N}\\
    \end{array}\right),
\end{align}
which implies that $\bar{\mathcal{J}}_{\rm rot}^{\rm N}$ does not contribute to $\delta\boldsymbol{p}$. 

Whereas $\bar{\mathcal{J}}_{\rm rot}^{\rm N}$ does not contribute to $\delta\boldsymbol{p}$, $\bar{\mathcal{J}}_{X}^{\rm N}$ and $\bar{\mathcal{J}}_{Y}^{\rm N}$ and $\bar{\mathcal{J}}_{XY}^{\rm N}$ contribute to the time evolution $\delta\boldsymbol{p}$.
In Eqs.~\eqref{ES11} and \eqref{ES12}, we obtain
\begin{align}
    \bar{\mathcal{J}}_{X}^{\rm N}&=\frac{\delta \mathbb{P}_{X}}{ \tau },
    \end{align}
    \begin{align}
    \bar{\mathcal{J}}_{Y}^{\rm N}&=\frac{\delta \mathbb{P}_{Y}}{ \tau },
\end{align}
if we define total probability flows in the marginal distributions $\delta\mathbb{P}_{X}:=\sum_y\delta p_{(1,y)}=\delta p_{v_2}+\delta p_{v_3}$, and $\delta\mathbb{P}_{Y}:=\sum_x\delta p_{(x,1)}=\delta p_{v_3}+\delta p_{v_4}$. Thus, $\bar{\mathcal{J}}_{X}^{\rm N}$ and $\bar{\mathcal{J}}_{Y}^{\rm N}$ correspond to the probability flows in the marginal distributions of subsystem $X$ and $Y$, respectively. Here, $\bar{\mathcal{J}}_{XY}^{\rm N}$ represents probability flows concerning the interaction between the subsystems, which cannot be written only by the contribution of $\bar{\mathcal{J}}_{X}^{\rm N}$ and $\bar{\mathcal{J}}_{Y}^{\rm N}$. 

\section{Calculation of partial entropy productions in Nash equilibrium}
In this section, we calculate $\Sigma_X^{\rm N}$ and $\Sigma_Y^{\rm N}$ in the limits $\lambda_X \to \infty$ with the fixed ratio $r= \lambda_X/ \lambda_Y$ and derive Eqs.~\eqref{E12} and \eqref{E14}. To calculate $\Sigma_X^{\rm N}$, we use the following notations
\begin{align}
    k_1&=-\alpha_{e_4}\delta p_{v_1}+\alpha_{e_2}\delta p_{v_2},
    \label{ES20}\\
    k_3&=-\alpha_{e_2}\delta p_{v_3}+\alpha_{e_4}\delta p_{v_4}.
    \label{ES21}
\end{align}
By using these notations, we obtain 
\begin{align}
    k_1-k_3&=(-\alpha_{e_4}\delta p_{v_1}+\alpha_{e_2}\delta p_{v_2})-(-\alpha_{e_2}\delta p_{v_3}+\alpha_{e_4}\delta p_{v_4})    \nonumber\\
    &=\alpha_{e_2}(\delta p_{v_2}+\delta p_{v_3})-\alpha_{e_4}(\delta p_{v_4}+\delta p_{v_1})
    \nonumber\\
    &=(\alpha_{e_2} +\alpha_{e_4})(\delta p_{v_2}+\delta p_{v_3})-\alpha_{e_4} \left(\sum_i \delta p_{v_i} \right) \nonumber \\
    &= \alpha_{Y} \tau \bar{\cal J}_{X}^{\rm N},
    \label{ES22}
\end{align}
where we used $\bar{\cal J}_{X}^{\rm N} = (\delta p_{v_2}+\delta p_{v_3})/\tau $, $\alpha_{Y}= \alpha_{e_2} + \alpha_{e_4}$, and $\sum_i \delta p_{v_i} =0$. We also obtain
\begin{align} 
    \alpha_{e_1}k_1+\alpha_{e_3}k_3&=\alpha_{e_1}(-\alpha_{e_4}\delta p_{v_1}+\alpha_{e_2}\delta p_{v_2})+\alpha_{e_3}(-\alpha_{e_2}\delta p_{v_3}+\alpha_{e_4}\delta p_{v_4})
    \nonumber\\
    &=-\alpha_{e_4}\alpha_{e_1}\delta p_{v_1}+\alpha_{e_1}\alpha_{e_2}\delta p_{v_2}-\alpha_{e_2}\alpha_{e_3}\delta p_{v_3}+\alpha_{e_3}\alpha_{e_4}\delta p_{v_4}
    \nonumber\\
    &=  \alpha_{X}\alpha_{Y}\tau \bar{\cal J}_{XY}^{\rm N},
    \label{ES23}
\end{align}
where we used Eq.~\eqref{ES13}. By using these equations, $\bar{\cal J}_{X}^{\rm N} = (\delta p_{v_2}+\delta p_{v_3})/\tau$, $\sum_i \delta p_{v_i} = 0$, $\alpha_X = \alpha_{e_1}+\alpha_{e_3}$, and Eqs.~\eqref{ES07} and \eqref{ES09}, we calculate $\Sigma_X^{\rm N}$ as follows,
\begin{align}
    \Sigma_X^{\rm N}
    &=\tau \left\{\alpha_{e_1}(\bar{J}_{e_1}^{\rm N})^2+\alpha_{e_3}(\bar{J}_{e_3}^{\rm N})^2\right\}
    \nonumber\\
    &=\frac{\alpha_{e_1}\{\alpha_{e_3}(\delta p_{v_2}+\delta p_{v_3})+r(-\alpha_{e_4}\delta p_{v_1}+\alpha_{e_2}\delta p_{v_2})\}^2+\alpha_{e_3}\{\alpha_{e_1}(\delta p_{v_1}+\delta p_{v_4})+r(-\alpha_{e_2}\delta p_{v_3}+\alpha_{e_4}\delta p_{v_4})\}^2}{ \tau (\alpha_{X}+r\alpha_{Y})^2}
    \nonumber\\
    &=\frac{  \alpha_{e_1}\{\tau  \alpha_{e_3}\bar{\cal J}_{X}^{\rm N}+rk_1\}^2+\alpha_{e_3} \left\{-\tau \alpha_{e_1}\bar{\cal J}_{X}^{\rm N}+rk_3 + \alpha_{e_1} \left(\sum_i \delta p_{v_i}  \right) \right\}^2}{\tau (\alpha_{X}+r\alpha_{Y})^2}
    \nonumber\\
    &=\frac{\alpha_{e_1}\alpha_{e_3}\alpha_{X}\tau^2(\bar{\cal J}_{X}^{\rm N})^2+2\alpha_{e_1}\alpha_{e_3}\tau \bar{\cal J}_{X}^{\rm N}r(k_1-k_3)+r^2\{\alpha_{e_1}k_1^2+\alpha_{e_3}k_3^2\}}{\tau (\alpha_{X}+r\alpha_{Y})^2}
    \nonumber\\
    &=\frac{\alpha_{e_1}\alpha_{e_3}\alpha_{X}\tau^2(\bar{\cal J}_{X}^{\rm N})^2+2\alpha_{e_1}\alpha_{e_3}r\alpha_{Y}\tau^2(\bar{\cal J}_{X}^{\rm N})^2+r^2\alpha_{X}^{-1}\{\alpha_{e_1}\alpha_{e_3}(k_1^2+k_3^2)+\alpha_{e_1}^2k_1^2+\alpha_{e_3}^2k_3^2\}}{\tau(\alpha_{X}+r\alpha_{Y})^2}
    \nonumber\\
    &=\frac{\alpha_{e_1}\alpha_{e_3}\alpha_{X}\tau^2(\bar{\cal J}_{X}^{\rm N})^2+2\alpha_{e_1}\alpha_{e_3}r\alpha_{Y}\tau^2(\bar{\cal J}_{X}^{\rm N})^2+r^2\alpha_{X}^{-1}\{\alpha_{e_1}\alpha_{e_3}(k_1-k_3)^2+(\alpha_{e_1}k_1+\alpha_{e_3}k_3)^2\}}{\tau (\alpha_{X}+r\alpha_{Y})^2}
    \nonumber\\
    &=\frac{\alpha_{e_1}\alpha_{e_3}\alpha_{X} \tau^2(\bar{\cal J}_{X}^{\rm N})^2+2\alpha_{e_1}\alpha_{e_3}r\alpha_{Y} \tau^2(\bar{\cal J}_{X}^{\rm N})^2+r^2\alpha_{X}^{-1}\{ \alpha_{e_1}\alpha_{e_3} \alpha_{Y}^2 \tau^2(\bar{\cal J}_{X}^{\rm N})^2+\alpha_{X}^2\alpha_{Y}^2\tau^2(\bar{\cal J}_{XY}^{\rm N})^2\}}{\tau (\alpha_{X}+r\alpha_{Y})^2}
    \nonumber\\
    &=\tau \left\{ \frac{\alpha_{e_1}\alpha_{e_3}\alpha_{X}^{-1} (\alpha_{X}^2 + 2 r \alpha_X \alpha_Y + r^2 \alpha_Y)^2(\bar{\cal J}_{X}^{\rm N})^2 }{ (\alpha_{X}+r\alpha_{Y})^2} +\frac{r^2\alpha_{X}\alpha_{Y}^2}{(\alpha_{X}+r\alpha_{Y})^2}(\bar{\cal J}_{XY}^{\rm N})^2\right\}
    \nonumber\\
    &=\tau \left\{\frac{1}{\sum_{e_i\in{\cal E}_X}\alpha_{e_i}^{-1}}(\bar{\cal J}_{X}^{\rm N})^2+\frac{r^2\alpha_{X}\alpha_{Y}^2}{(\alpha_{X}+r\alpha_{Y})^2}(\bar{\cal J}_{XY}^{\rm N})^2\right\}. 
    \label{ES24}
\end{align}
By using $\gamma= \alpha_X/ \alpha_Y$, $f_X(r;\gamma) = r^2 (\gamma +1)/(\gamma + r)^2$, $\Sigma_{X}^{\rm min} = (\int^{t_0+\tau}_{t_0} dt \bar{\mathcal{J}}^{\rm N}_X)^2/ [\tau (\sum_{e_i\in{\cal E}_X} \alpha_{e_i}^{-1})] = \tau  (\bar{\mathcal{J}}^{\rm N}_X)^2/ (\sum_{e_i\in{\cal E}_X} \alpha_{e_i}^{-1})$ and $\Sigma_{XY} =  (\int^{t_0+\tau}_{t_0} dt \bar{\mathcal{J}}^{\rm N}_{XY})^2/ [\tau (\alpha_{X}^{-1}+ \alpha_{Y}^{-1})] = \tau  (\bar{\mathcal{J}}^{\rm N}_{XY})^2/ (\alpha_{X}^{-1}+ \alpha_{Y}^{-1})$, we obtain Eq.~\eqref{E12},
\begin{align}
    \Sigma_X^{\rm N}
    &=\tau \left\{\frac{1}{\sum_{e_i\in{\cal E}_X}\alpha_{e_i}^{-1}}(\bar{\cal J}_{X}^{\rm N})^2+\frac{r^2\alpha_{X}\alpha_{Y}^2}{(\alpha_{X}+r\alpha_{Y})^2}(\bar{\cal J}_{XY}^{\rm N})^2\right\}
    \nonumber\\
    &=\Sigma_{X}^{\rm min}+ \frac{r^2 (\alpha_{X}\alpha_{Y} +\alpha_{Y}^2 ) }{(\alpha_{X}+r\alpha_{Y})^2} \Sigma_{XY} \nonumber\\
    &=\Sigma_{X}^{\rm min}+ \frac{r^2 (\gamma +1) }{(\gamma+r)^2} \Sigma_{XY} \nonumber\\
    &=\Sigma_{X}^{\rm min}+f_X(r;\gamma)\Sigma_{XY}.
\end{align}

To calculate $\Sigma_Y^{\rm N}$, we use the following notations
\begin{align}
    k_2&=-\alpha_{e_1}\delta p_{v_2}+\alpha_{e_3}\delta p_{v_3},
    \label{ES25}\\
    k_4&=-\alpha_{e_3}\delta p_{v_4}+\alpha_{e_1}\delta p_{v_1}.
    \label{ES26}
\end{align}
By using these notations, we obtain
\begin{align}
    k_2-k_4&=(-\alpha_{e_1}\delta p_{v_2}+\alpha_{e_3}\delta p_{v_3})-(-\alpha_{e_3}\delta p_{v_4}+\alpha_{e_1}\delta p_{v_1})
    \nonumber\\
    &=\alpha_{e_3}(\delta p_{v_3}+\delta p_{v_4})-\alpha_{e_1}(\delta p_{v_1}+\delta p_{v_2})
    \nonumber\\
    &=(\alpha_{e_1} +\alpha_{e_3})(\delta p_{v_3}+\delta p_{v_4})-\alpha_{e_1} \left(\sum_i \delta p_{v_i} \right) \nonumber \\
    &=\alpha_{X}\tau \bar{\cal J}_{Y}^{\rm N},
    \label{ES27}
\end{align}
where we used $\bar{\cal J}_{Y}^{\rm N} = (\delta p_{v_3}+\delta p_{v_4})/\tau $, $\alpha_{X}= \alpha_{e_1} + \alpha_{e_3}$ and $\sum_i \delta p_{v_i} =0$. We also obtain
\begin{align}
    \alpha_{e_2}k_2+\alpha_{e_4}k_4&=\alpha_{e_2}(-\alpha_{e_1}\delta p_{v_2}+\alpha_{e_3}\delta p_{v_3})+\alpha_{e_4}(-\alpha_{e_3}\delta p_{v_4}+\alpha_{e_1}\delta p_{v_1})
    \nonumber\\
    &=-(-\alpha_{e_4}\alpha_{e_1}\delta p_{v_1}+\alpha_{e_1}\alpha_{e_2}\delta p_{v_2}-\alpha_{e_2}\alpha_{e_3}\delta p_{v_3}+\alpha_{e_3}\alpha_{e_4}\delta p_{v_4} )
    \nonumber\\
    &=- \alpha_{X}\alpha_{Y}\tau \bar{\cal J}_{XY}^{\rm N}, 
    \label{ES28}
\end{align}
where we used Eq.~\eqref{ES13}. By using these equations, $\bar{\cal J}_{Y}^{\rm N} = (\delta p_{v_3}+\delta p_{v_4})/\tau$, $\sum_i \delta p_{v_i} = 0$, $\alpha_Y = \alpha_{e_2}+\alpha_{e_4} $, and Eqs.~\eqref{ES08} and \eqref{ES10}, we calculate $\Sigma_Y^{\rm N}$ as follows,
\begin{align}
    \Sigma_Y^{\rm N} &=\tau \left\{\alpha_{e_2}(\bar{J}_{e_2}^{\rm N})^2+\alpha_{e_4}(\bar{J}_{e_4}^{\rm N})^2\right\}
    \nonumber\\
    &=\frac{\alpha_{e_2}\{r\alpha_{e_4}(\delta p_{v_3}+\delta p_{v_4})-\alpha_{e_1}\delta p_{v_2}+\alpha_{e_3}\delta p_{v_3}\}^2+\alpha_{e_4}\{r\alpha_{e_2}(\delta p_{v_1}+\delta p_{v_2})-\alpha_{e_3}\delta p_{v_4}+\alpha_{e_1}\delta p_{v_1}\}^2}{\tau (\alpha_{X}+r\alpha_{Y})^2}
    \nonumber\\
    &=\frac{\alpha_{e_2}\{r\alpha_{e_4} \tau \bar{\cal J}_{Y}^{\rm N}+k_2\}^2+\alpha_{e_4} \left\{-r\alpha_{e_2}\tau \bar{\cal J}_{Y}^{\rm N}+k_4 + r\alpha_{e_2} \left( \sum_i \delta p_{v_i}\right) \right\}^2}{\tau (\alpha_{X}+r\alpha_{Y})^2}
    \nonumber\\
    &=\frac{\alpha_{e_2}\alpha_{e_4}r^2\alpha_{Y}\tau^2(\bar{\cal J}_{Y}^{\rm N})^2+2\alpha_{e_2}\alpha_{e_4}\tau \bar{\cal J}_{Y}^{\rm N}r(k_2-k_4)+\{\alpha_{e_2}k_2^2+\alpha_{e_4}k_4^2\}}{\tau (\alpha_{X}+r\alpha_{Y})^2}
    \nonumber\\
    &=\frac{\alpha_{e_2}\alpha_{e_4}r^2\alpha_{Y}\tau^2(\bar{\cal J}_{Y}^{\rm N})^2+2\alpha_{e_2}\alpha_{e_4}r\alpha_{X}\tau^2(\bar{\cal J}_{Y}^{\rm N})^2+\alpha_{Y}^{-1}\{\alpha_{e_2}\alpha_{e_4}(k_2^2+k_4^2)+\alpha_{e_2}^2k_2^2+\alpha_{e_4}^2k_4^2\}}{\tau (\alpha_{X}+r\alpha_{Y})^2}
    \nonumber\\
    &=\frac{\alpha_{e_2}\alpha_{e_4}r^2\alpha_{Y}\tau^2(\bar{\cal J}_{Y}^{\rm N})^2+2\alpha_{e_2}\alpha_{e_4}r\alpha_{X}\tau^2(\bar{\cal J}_{Y}^{\rm N})^2+\alpha_{Y}^{-1}\{\alpha_{e_2}\alpha_{e_4}(k_2-k_4)^2+(\alpha_{e_2}k_2+\alpha_{e_4}k_4)^2\}}{\tau (\alpha_{X}+r\alpha_{Y})^2}
    \nonumber\\
    &=\frac{\alpha_{e_2}\alpha_{e_4}r^2\alpha_{Y}\tau^2(\bar{\cal J}_{Y}^{\rm N})^2+2\alpha_{e_2}\alpha_{e_4}r\alpha_{X}\tau^2(\bar{\cal J}_{Y}^{\rm N})^2+\alpha_{Y}^{-1}\{\alpha_{e_2}\alpha_{e_4} \alpha_{X}^2\tau^2(\bar{\cal J}_{Y}^{\rm N})^2+\alpha_{X}^2\alpha_{Y}^2\tau^2(\bar{\cal J}_{XY}^{\rm N})^2\}}{\tau (\alpha_{X}+r\alpha_{Y})^2}
    \nonumber\\
    &=\tau \left\{ \frac{\alpha_{e_2}\alpha_{e_4}\alpha_{Y}^{-1} (r^2\alpha_{Y}^2 + 2 r \alpha_X \alpha_Y +  \alpha_X)^2(\bar{\cal J}_{Y}^{\rm N})^2 }{ (\alpha_{X}+r\alpha_{Y})^2} +\frac{\alpha_{X}^2\alpha_{Y}}{(\alpha_{X}+r\alpha_{Y})^2}(\bar{\cal J}_{XY}^{\rm N})^2\right\}
    \nonumber\\
    &=\tau \left\{\frac{1}{\sum_{e_i\in{\cal E}_Y}\alpha_{e_i}^{-1}}(\bar{\cal J}_{Y}^{\rm N})^2+\frac{\alpha_{X}^2\alpha_{Y}}{(\alpha_{X}+r\alpha_{Y})^2}(\bar{\cal J}_{XY}^{\rm N})^2\right\}.
    \label{ES29}
\end{align}
By using $\gamma= \alpha_X/ \alpha_Y$, $f_Y(r;\gamma) =  \gamma (\gamma +1)/(\gamma + r)^2$, $\Sigma_{Y}^{\rm min} = (\int^{t_0+\tau}_{t_0} dt \bar{\mathcal{J}}^{\rm N}_Y)^2/ [\tau (\sum_{e_i\in{\cal E}_Y} \alpha_{e_i}^{-1})] = \tau  (\bar{\mathcal{J}}^{\rm N}_Y)^2/ (\sum_{e_i\in{\cal E}_Y} \alpha_{e_i}^{-1})$ and $\Sigma_{XY} = \tau  (\bar{\mathcal{J}}^{\rm N}_{XY})^2/ (\alpha_{X}^{-1}+ \alpha_{Y}^{-1})$, we obtain Eq.~\eqref{E14},
\begin{align}
    \Sigma_Y^{\rm N}
    &=\tau \left\{\frac{1}{\sum_{e_i\in{\cal E}_Y}\alpha_{e_i}^{-1}}(\bar{\cal J}_{Y}^{\rm N})^2+\frac{\alpha_{X}^2\alpha_{Y}}{(\alpha_{X}+r\alpha_{Y})^2}(\bar{\cal J}_{XY}^{\rm N})^2\right\}
    \nonumber\\
    &=\Sigma_{Y}^{\rm min}+ \frac{ (\alpha_{X}\alpha_{Y} +\alpha_{X}^2 ) }{(\alpha_{X}+r\alpha_{Y})^2} \Sigma_{XY} \nonumber\\
    &=\Sigma_{X}^{\rm min}+ \frac{\gamma (\gamma +1) }{(\gamma+r)^2} \Sigma_{XY} \nonumber\\
    &=\Sigma_{Y}^{\rm min}+f_Y(r;\gamma)\Sigma_{XY}.
\end{align}

\section{Relation between optimal transport theory and the main result}
We here remark on a relation between optimal transport theory and the main result in the letter. 

At first, we explained the conventional optimal transport theory~\cite{villani2021topics} for the continuous state. The optimal transport theory for the continuous state has been discussed in terms of the minimization of the entropy production for the Fokker--Planck equation,
\begin{align}
    \partial_t P^{(t)}(\boldsymbol{x} ) = - \nabla \cdot (\boldsymbol{\nu}^{(t)}(\boldsymbol{x})  P^{(t)}(\boldsymbol{x} )), \\
    \boldsymbol{\nu}^{(t)}(\boldsymbol{x}) = \mu F^{(t)} - T \nabla \ln P^{(t)}(\boldsymbol{x} ),
\end{align}
where $\boldsymbol{x} \in \mathbb{R}^d$ is a $d$-dimensional continuous state, $P^{(t)}(\boldsymbol{x} ) $ is the probability density that satisfies $\int d\boldsymbol{x} P^{(t)}(\boldsymbol{x} ) =1$ and $ P^{(t)}(\boldsymbol{x} ) \geq 0$, $T$ is the temperature and $\mu$ is the mobility. The entropy production is obtained as the quadratic functions,
\begin{align}
    \Sigma_{\rm tot} =\int_{t_0}^{t_1} \int d\boldsymbol{x} \boldsymbol{\nu}^{(t)}(\boldsymbol{x}) \cdot \boldsymbol{\nu}^{(t)}(\boldsymbol{x}) \frac{  P^{(t)}(\boldsymbol{x} ) }{\mu T}, 
\end{align}
and its minimization for the fixed initial and final states $P^{(t_0)}(\boldsymbol{x} ) $ and $P^{(t_1)}(\boldsymbol{x} )$ is given by
\begin{align}
    \Sigma_{\rm tot}  \geq {\rm min}_{(\boldsymbol{\nu}^{*(t)})_{t_0 \leq t \leq t_1}}\int_{t_0}^{t_1} dt \int d\boldsymbol{x} \frac{ (\boldsymbol{\nu}^{*(t)}(\boldsymbol{x}) P^{(t)}(\boldsymbol{x} )) \cdot (\boldsymbol{\nu}^{*(t)}(\boldsymbol{x}) P^{(t)}(\boldsymbol{x} ) ) }{P^{(t)}(\boldsymbol{x} ) \mu T}, 
\end{align}
where $(\boldsymbol{\nu}^{*(t)})_{t_0 \leq t \leq t_1}$ provides the same time evolution for $(\boldsymbol{\nu}^{(t)})_{t_0 \leq t \leq t_1}$,
\begin{align}
 \partial_t P^{(t)}(\boldsymbol{x} )   = - \nabla \cdot ( \boldsymbol{\nu}^{(t)}(\boldsymbol{x}) P^{(t)}(\boldsymbol{x} )) = - \nabla \cdot ( \boldsymbol{\nu}^{*(t)}(\boldsymbol{x}) P^{(t)}(\boldsymbol{x} )).
\end{align}
Here, we only assume that $(\boldsymbol{\nu}^{(t)})_{t_0 \leq t \leq t_1}$ satisfies the boundary conditions $P^{(t_0)}(\boldsymbol{x} ) $ and $P^{(t_1)}(\boldsymbol{x} )$ with the continuity equation $\partial_t P^{(t)}(\boldsymbol{x} )  = - \nabla \cdot ( \boldsymbol{\nu}^{(t)}(\boldsymbol{x}) P^{(t)}(\boldsymbol{x} )) $.
In optimal transport theory, this minimization is given by the $L^2$-Wasserstein distance $\mathcal{W}(P^{(0)}, P^{(\tau)})$~\cite{benamou2000computational} defined as
\begin{align}
     \mathcal{W}(P^{(0)}, P^{(\tau)}) = \sqrt{ {\rm min}_{(\boldsymbol{\nu}^{*(t)})_{t_0 \leq t \leq t_1}} \tau \int_{t_0}^{t_0+\tau} dt \int d\boldsymbol{x} \boldsymbol{\nu}^{*(t)}(\boldsymbol{x}) \cdot \boldsymbol{\nu}^{*(t)}(\boldsymbol{x})  P^{(t)}(\boldsymbol{x})} ,
\end{align}
thus the lower bound on the entropy production~\cite{aurell2012refined} is obtained as
\begin{align}
   \Sigma_{\rm tot}  \geq  \frac{[\mathcal{W}(P^{(0)}, P^{(\tau)})]^2}{\mu T\tau}.
\end{align}
This result is also recently discussed in terms of the thermodynamic speed limit~\cite{nakazato2021geometrical, dechant2022geometric}. The thermodynamic speed limit provides the lower bound on the entropy production which is proportional to $1/\tau$. In Ref.~\cite{nakazato2021geometrical}, the thermodynamic speed limits for the partial entropy production $\Sigma_X$ and $\Sigma_Y$ are also obtained. 

To generalize the above fact for dynamics of the discrete state described by the Markov jump processes, J. Maas proposed a correspondence of the $L^2$-Wasserstein distance~\cite{maas2011gradient} for the Markov jump processes. We also discussed the minimum entropy production based on this correspondence of the $L^2$-Wasserstein distance for the Markov jump process~\cite{yoshimura2023housekeeping}. For general Markov jump networks, for example, a bipartite model in the letter, we can use the incidence matrix as follows,
\begin{align}
    \frac{d}{dt} \boldsymbol{p}^{(t)}&=\mathsf{B} \boldsymbol{J}^{(t)},
\end{align}
where $\mathsf{B}$ is the incidence matrix and $(\boldsymbol{J}^{(t)})_{\rho} = {J}^{(t)}_{\rho}= J^{+(t)}_{\rho} - J^{-(t)}_{\rho}$ is the vector of the flow on each edge, where $J^{+(t)}_{\rho} = W_{z'\rightarrow z}^{(t)}p_{z'}^{(t)}$ and $J^{-(t)}_{\rho} = W_{z\rightarrow z'}^{(t)}p_{z}^{(t)}$ for the directed edge $\rho = z'\rightarrow z$. We only assume that $d \boldsymbol{p}^{(t)}/dt=\mathsf{B} \boldsymbol{J}^{(t)}$ satisfies the boundary condition $\boldsymbol{p}^{(t_0)} = \boldsymbol{p}^{\rm i}$ and $\boldsymbol{p}^{(t_1)} = \boldsymbol{p}^{\rm f}$ as discussed in the letter. The entropy production is obtained as the product of the force $(\boldsymbol{F}^{(t)} )_{\rho} ={F}^{(t)}_{\rho}= \ln ( J^{+(t)}_{\rho}/J^{-(t)}_{\rho})$ and the flow ${J}^{(t)}_{\rho}$,
\begin{align}
     \Sigma_{\rm tot} = \int_{t_0}^{t_0 + \tau} dt  \sum_{\rho} {F}^{(t)}_{\rho} {J}^{(t)}_{\rho} =\int_{t_0}^{t_0 + \tau} dt   {J}^{(t)}_{\rho} \alpha_{\rho} {J}^{(t)}_{\rho}.
\end{align}
where $\alpha_{\rho} ={F}^{(t)}_{\rho} /{J}^{(t)}_{\rho} $ is (the reciprocal of) the edgewise Onsager coefficient that satisfies ${F}^{(t)}_{\rho} = \alpha_{\rho} {J}^{(t)}_{\rho}$. For the fixed $\alpha_{\rho}$, the minimum entropy production cannot be zero for the given transition from $\boldsymbol{p}^{(t_0)}$ to $\boldsymbol{p}^{(t_1)}$. This condition is satisfied automatically in the framework of linear irreversible thermodynamics. The minimum entropy production is given by
\begin{align}
     \Sigma_{\rm tot}  \geq \min_{({J}^{*(t)}_{\rho})_{t_0 \leq t \leq t_1}} \int_{t_0}^{t_0 + \tau} dt  \sum_{\rho} {J}^{*(t)}_{\rho} \alpha_{\rho} {J}^{*(t)}_{\rho}
\end{align}
where $({J}^{*(t)}_{\rho})_{t_0 \leq t \leq t_1}$ provides the same time evolution for  $({J}^{(t)}_{\rho})_{t_0 \leq t \leq t_1}$,
\begin{align}
    \frac{d}{dt} \boldsymbol{p}^{(t)}&=\mathsf{B} \boldsymbol{J}^{(t)} = \mathsf{B} \boldsymbol{J}^{*(t)}.
\end{align}
Here, we only assume that $(\boldsymbol{J}^{(t)})_{t_0 \leq t \leq t_1}$ satisfies the boundary conditions $\boldsymbol{p}^{(t_0)} = \boldsymbol{p}^{\rm i}$ and $\boldsymbol{p}^{(t_1)}=\boldsymbol{p}^{\rm f}$ with $d \boldsymbol{p}^{(t)}/dt =  {\sf B}  \boldsymbol{J}^{(t)}$.
If we compare this result with the result for the continuous case, we can find that the incidence matrix $\mathsf{B}$ corresponds to $- {\rm div} (\cdots)= - \nabla \cdot (\cdots)$, $\boldsymbol{J}^{(t)}$ corresponds to $\boldsymbol{\nu}^{(t)}(\boldsymbol{x})  P^{(t)}(\boldsymbol{x} )$ and $\alpha_{\rho}$ corresponds to $1/[\mu T P^{(t)} (\boldsymbol{x} )]$ respectively. In terms of the kernel, 
\begin{align}
  \mathsf{B} [\boldsymbol{J}^{(t)} - \boldsymbol{J}^{*(t)}] =\boldsymbol{0}
\end{align}
means $\boldsymbol{J}^{(t)} - \boldsymbol{J}^{*(t)} \in {\rm Ker} \mathsf{B} $. We also can define a correspondence of the $L^2$-Wasserstein distance by using 
\begin{align}
     \tilde{\mathcal{W}} (P^{(0)}, P^{(\tau)}) = \sqrt{\min_{{J}^{*(t)}_{\rho}} \tau  \int_{t_0}^{t_0 + \tau} dt \sum_{\rho} {J}^{*(t)}_{\rho} \alpha_{\rho} {J}^{*(t)}_{\rho} } ,
\end{align}
that provides the thermodynamic speed limit for the Markov jump networks~\cite{yoshimura2023housekeeping},
\begin{align}
      \Sigma  \geq  \frac{[\tilde{\mathcal{W}} (P^{(0)}, P^{(\tau)})]^2}{\tau}.
\end{align}

We here consider our result for the bipartite model in the letter from the viewpoint of optimal transport theory and the minimization of the entropy production. The concept of the rotation flow $\bar{\mathcal{J}}_{\rm rot}^{\rm N}$, which does not contribute to the time evolution of probability distribution, is related to the quantity $\boldsymbol{J}^{(t)} - \boldsymbol{J}^{*(t)}$ in the above discussion. Indeed, the probability flow $\bar{\mathcal{J}}_{\rm rot}^{\rm N}$ does not contribute the time evolution 
\begin{align}
\frac{d\boldsymbol{p}^{(t)}}{dt} = \mathsf{B} {\sf T}^{-1} \boldsymbol{\mathcal{J}}^{\rm N} = \mathsf{B} {\sf T}^{-1} (\bar{\mathcal{J}}_{X}^{\rm N},\bar{\mathcal{J}}_{Y}^{\rm N}, \bar{\mathcal{J}}_{XY}^{\rm N}, \bar{\mathcal{J}}_{\rm rot}^{\rm N})^{\rm T} = \mathsf{B} {\sf T}^{-1} (\bar{\mathcal{J}}_{X}^{\rm N},\bar{\mathcal{J}}_{Y}^{\rm N}, \bar{\mathcal{J}}_{XY}^{\rm N}, 0)^{\rm T},
\end{align}
 Thus, the contribution of the rotation flow ${\sf T}^{-1} (0,0,0, \bar{\mathcal{J}}_{\rm rot}^{\rm N})^{\rm T} \in {\rm Ker}\mathsf{B}$ corresponds to $\boldsymbol{J}^{(t)} - \boldsymbol{J}^{*(t)}$ in the optimal transport theory. This rotational flow $\bar{\mathcal{J}}_{\rm rot}^{\rm N}$ can be identified with the cycle flow for the bipartite system and the minimum entropy production can be achieved when $\boldsymbol{J}^{(t)} - \boldsymbol{J}^{*(t)}$ vanishes without any assumption in the theory~\cite{yoshimura2023housekeeping}.

In the letter, we consider a more complicated problem of the minimum entropy productions by focusing on the Nash equilibrium solution, which is not discussed in Ref.~\cite{yoshimura2023housekeeping}. Thus, the minimum entropy production is not necessarily achieved when $\boldsymbol{J}^{(t)} - \boldsymbol{J}^{*(t)}= \boldsymbol{0}$ or $\bar{\mathcal{J}}_{\rm rot}^{\rm N}=0$. However, the minimum total entropy production $\Sigma_{\rm tot}^{\rm min}$, and the minimum partial entropy productions $\Sigma_X^{\rm min}$ and $\Sigma_Y^{\rm min}$ can be achieved when the mode of rotation flow $\bar{\mathcal{J}}_{\rm rot}^{\rm N}$ is changed.

We also remark on the thermodynamic speed limit for partial entropy production. In the main result, we showed that
\begin{align}
\Sigma_X \geq \Sigma_X^{\rm min}=\frac{(\int_{t_0}^{t_0+\tau} dt \bar{\mathcal{J}}_{X}^{\rm N})^2}{\tau(\sum_{e_i\in{\cal E}_X}\alpha_{e_i}^{-1} )},
\end{align}
under the constraint on  $\min_{\hat{W}_{Y}} C_Y$, and 
\begin{align}
\Sigma_Y \geq \Sigma_Y^{\rm min}=\frac{(\int_{t_0}^{t_0+\tau} dt \bar{\mathcal{J}}_{Y}^{\rm N})^2}{\tau(\sum_{e_i\in{\cal E}_Y}\alpha_{e_i}^{-1} )}
\end{align}
under the constraint on  $\min_{\hat{W}_{X}} C_X$ when $\lambda_X{\to}\infty$ and $\lambda_Y{\to}\infty$ that means $\boldsymbol{p}^{(t_0)}=\boldsymbol{p}^{\rm i}$ and $\boldsymbol{p}^{(t_1)}=\boldsymbol{p}^{\rm f}$.
Here, the time evolution of the marginal distributions, defined as  $\mathbb{P}_{X}^{(t)}:=\sum_yp^{(t)} _{(1,y)}=p^{(t)} _{v_2}+p^{(t)} _{v_3}$, and $\mathbb{P}_{Y}^{(t)}:=\sum_x p^{(t)} _{(x,1)}=p^{(t)} _{v_3}+p^{(t)} _{v_4}$, are given by 
\begin{align}
    &\frac{d}{dt}\left(\begin{array}{c}
        \mathbb{P}_{X}^{(t)}  \\
        \mathbb{P}_{Y}^{(t)}  
    \end{array}\right) =\left(\begin{array}{c}
        \mathcal{J}_{X}^{\rm N} \\
        \mathcal{J}_{Y}^{\rm N}
    \end{array}\right).
    \label{E44}
\end{align}
Thus, the terms $(\int_{t_0}^{t_0+\tau} dt \bar{\mathcal{J}}_{X}^{\rm N})^2$ and $(\int_{t_0}^{t_0+\tau} dt \bar{\mathcal{J}}_{Y}^{\rm N})^2$ can be interpreted as
\begin{align}
    \left(\int_{t_0}^{t_0+\tau} dt \bar{\mathcal{J}}_{X}^{\rm N} \right)^2 =& |\mathbb{P}_{X}^{(t_0)}  - \mathbb{P}_{X}^{(t_0+ \tau)} |^2, \\
     \left(\int_{t_0}^{t_0+\tau} dt \bar{\mathcal{J}}_{Y}^{\rm N} \right)^2 =& |\mathbb{P}_{Y}^{(t_0)}  - \mathbb{P}_{Y}^{(t_0+ \tau)} |^2.
\end{align}
Finally, we obtain the lower bounds on the partial entropy productions
\begin{align}
    &\Sigma_X \geq \frac{|\mathbb{P}_{X}^{(t_0)}  - \mathbb{P}_{X}^{(t_0+ \tau)} |^2 }{\tau(\sum_{e_i\in{\cal E}_X}\alpha_{e_i}^{-1} )} ,\\
    &\Sigma_Y \geq \frac{|\mathbb{P}_{Y}^{(t_0)}  - \mathbb{P}_{Y}^{(t_0+ \tau)} |^2 }{\tau(\sum_{e_i\in{\cal E}_Y}\alpha_{e_i}^{-1} )},
\end{align}
which are regarded as thermodynamic speed limits for the partial entropy productions $\Sigma_X$ and $\Sigma_Y$ under the constraint on $\min_{\hat{W}_{Y}} C_Y$ and $\min_{\hat{W}_{X}} C_X$, respectively.

\section{Nash equilibrium solution on finite weights of error penalty}
In the letter, we considered the Nash equilibrium solution and partial entropy production on infinite weights of error penalty, i.e., $\lambda_X\to \infty$ with the fixed ratio $r$, where the final state completely matches the target state $\boldsymbol{p}^{(t_1)}=\boldsymbol{p}^{\rm f}$. In this section, we consider cases of finite values of $\lambda_X$ and $\lambda_Y$.

Fig.~E\ref{F03} considers the Nash equilibrium when $\lambda_X$ is fixed to a finite value, whereas $\lambda_Y$ varies. Fig.~E\ref{F03}-(a) shows probability flows $\bar{\boldsymbol{\mathcal{J}}}^{\rm N}$. As different to the case in $\lambda_X\to \infty$ with the fixed ratio $r$, $\bar{\mathcal{J}}_{X}^{\rm N}$, $\bar{\mathcal{J}}_{Y}^{\rm N}$, and $\bar{\mathcal{J}}_{XY}^{\rm N}$ are not invariant on $r$. Fig.~E\ref{F03}-(b) shows the entropy productions of $\Sigma_X^{\rm N}$, $\Sigma_Y^{\rm N}$, and $\Sigma_{\rm tot}^{\rm N}$. We can also see the trade-off relation between partial entropy productions for the finite values of $\lambda_X$ and $\lambda_Y$. The minimum partial entropy production in $X$ is achieved when $r \to 0$ and the minimum partial entropy production in $Y$ is achieved when $r \to \infty$. However, the minimum total entropy production is not always achieved when $r=1$.
\begin{figure}[htbp]
\begin{center}
\includegraphics[width=0.65\linewidth]{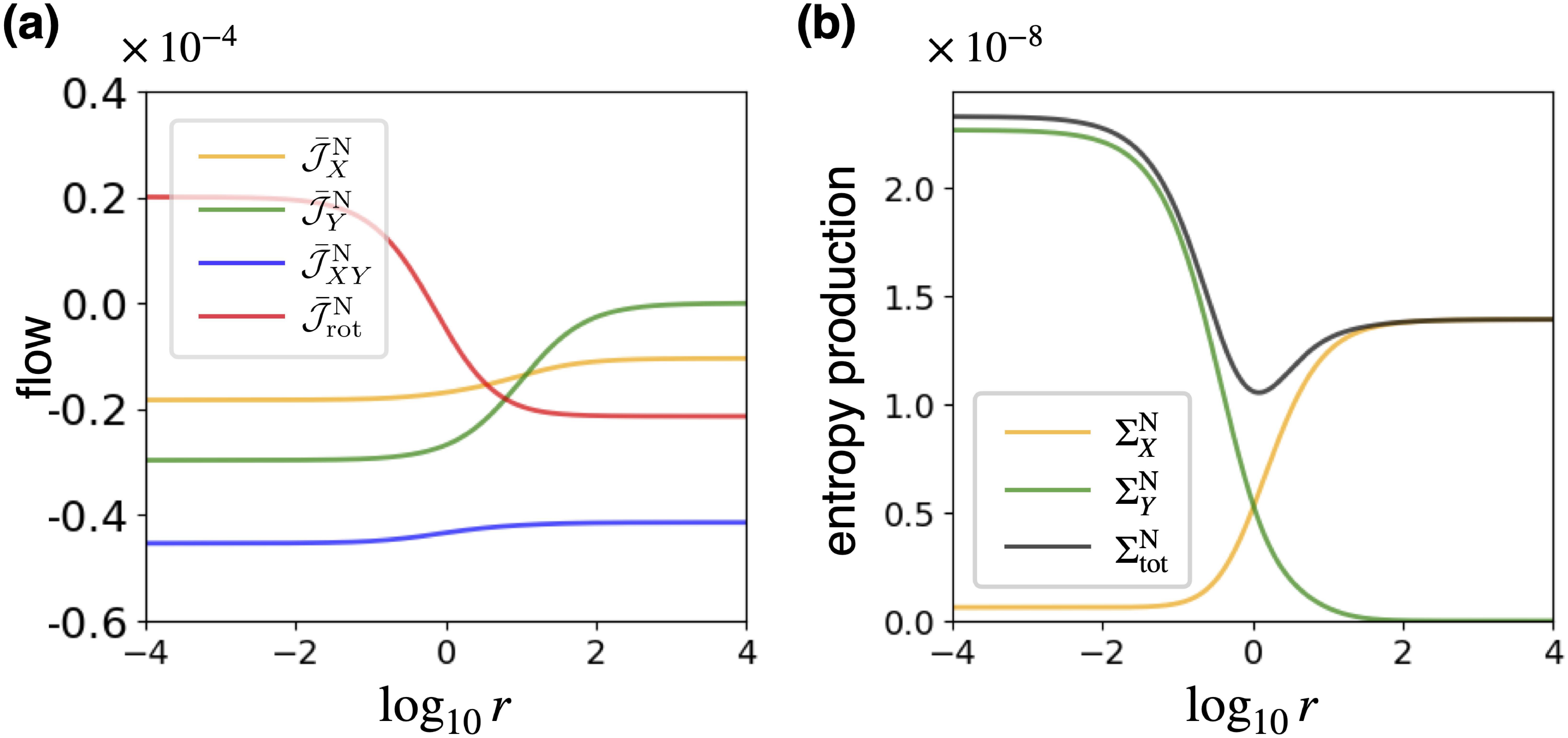}
\caption{(a) Probability flows and (b) entropy productions in $\lambda_X=10$ and $10^{-4}\le r\le 10^4$. The other various parameters are equal to those in Fig.~\ref{F02}. The horizontal axis indicates $\log r$ again. In panel (a), the orange, green, blue, and red lines indicate $\bar{\mathcal{J}}_{X}^{\rm N}$, $\bar{\mathcal{J}}_{Y}^{\rm N}$, $\bar{\mathcal{J}}_{XY}^{\rm N}$, and $\bar{\mathcal{J}}_{\rm rot}^{\rm N}$, respectively. In panel (b), the orange, green, and black lines indicate $\Sigma_X^{\rm N}$, $\Sigma_Y^{\rm N}$, and $\Sigma_{\rm tot}^{\rm N}$, respectively.}
\label{F03}
\end{center}
\end{figure}

Fig.~E\ref{F04} shows the dependence of $\Sigma_X^{\rm N}$ on $\lambda_X$. From this figure, we see that $\Sigma_X^{\rm N}$ is almost zero in the limit $\lambda_X \to 0$. The partial entropy production $\Sigma_X^{\rm N}$ monotonically increases as $\lambda_X$ becomes larger. This monotonic behavior implies the effect of the penalty of a given state transition in the minimization problem of the partial entropy productions. If the penalty is smaller, the partial entropy production can be minimized much more. In the limit $\lambda_X\to 0$, a state transition in the subsystem $X$ does not occur to maintain the partial entropy production to be zero.
\begin{figure}[htbp]
\begin{center}
\includegraphics[width=0.35\linewidth]{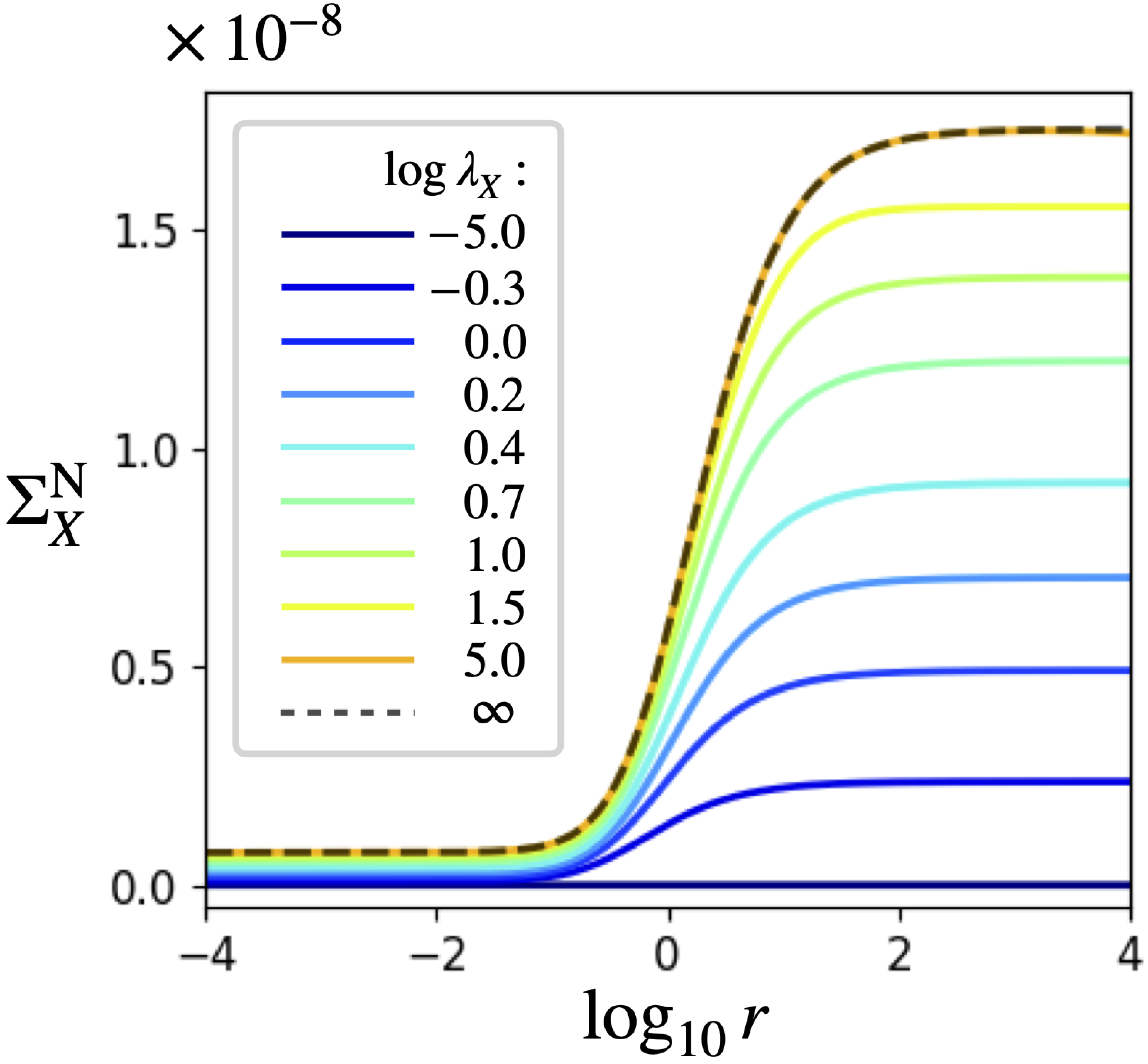}
\caption{Dependence of $\Sigma_X^{\rm N}$ on $\lambda_X$. The various parameters other than $\lambda_X$ are equal to those in Fig.~\ref{F02}. The horizontal axis is $\log r$ again. The color difference represents the difference in $\lambda_X$, and the closer it gets from blue to orange, the larger $\lambda_X$ is. In final, the line for $\lambda_X=10^{15}$ almost matches the limit value of $\lambda_X\rightarrow\infty$ represented by the dashed line in black.}
\label{F04}
\end{center}
\end{figure}

\end{widetext}


%

\end{document}